\documentclass[lettersize,journal]{IEEEtran}
\usepackage{amsmath,amsfonts}
\usepackage{algorithmic}
\usepackage{algorithm}
\usepackage{array}
\usepackage[caption=false,font=footnotesize,labelfont=sf,textfont=sf]{subfig}
\usepackage{textcomp}
\usepackage{stfloats}
\usepackage{url}
\usepackage{verbatim}
\usepackage{graphicx}
\usepackage{cite}
\usepackage{amsmath,amsfonts}
\usepackage{algorithmic}
\usepackage{algorithm}
\usepackage{array}
\usepackage{textcomp}
\usepackage{stfloats}
\usepackage{url}
\usepackage{verbatim}
\usepackage{graphicx}
\usepackage{cite}
\usepackage{romannum}
\usepackage{tabularx}
\usepackage{multirow}
\usepackage{changepage} % For adjustwidth environment
\usepackage{threeparttable} % For captions that span full width
\usepackage{pgfplots}
\usepackage{orcidlink} 
\usepackage{tikz}
\usepackage{booktabs}
\usepackage{multirow}
\usepackage{xcolor}
\usepackage{pgfplots}
\usepgfplotslibrary{groupplots}
\pgfplotsset{compat=1.18}
\usepackage[most]{tcolorbox}
\usepackage{ragged2e}
\usepackage{fontawesome5}
\usepackage{fvextra}
\usepackage{graphicx}
\usepackage{subcaption}
\usepackage[caption=false,font=footnotesize]{subfig}
\definecolor{contextgray}{HTML}{F1F1F1}
\definecolor{audioblue}{HTML}{DDEAF7}
\definecolor{audiotitle}{HTML}{2F69A1}
\definecolor{textyellow}{HTML}{FFF1C9}
\definecolor{texttitle}{HTML}{9A7200}
\definecolor{reasonred}{HTML}{FADFDB}
\definecolor{reasontitle}{HTML}{A93226}
\definecolor{correctgreen}{HTML}{2E7D32}
\definecolor{wrongred}{HTML}{D64B3C}
\definecolor{cardborder}{HTML}{D8D8D8}

\usepackage[most]{tcolorbox}
\usepackage{xcolor}

\definecolor{sftbg}{HTML}{F3F6FA}
\definecolor{grpobg}{HTML}{E1EAF6}
\definecolor{oursbg}{HTML}{C9DAEF}

\definecolor{contextbg}{HTML}{F0F0F0}
\definecolor{errorred}{HTML}{B02418}
\definecolor{evidencegreen}{HTML}{4C7C31}
\definecolor{bodyblack}{HTML}{171717}

\newtcolorbox{reasoningcard}[2][]{%
    enhanced,
    colback=#2,
    colframe=#2,
    boxrule=0pt,
    arc=1.6mm,
    outer arc=1.6mm,
    boxsep=0pt,
    left=2mm,
    right=2mm,
    top=1mm,
    bottom=1mm,
    fontupper=\footnotesize,
    before upper={%
        \raggedright
        \color{bodyblack}
        \linespread{0.96}\selectfont
        \setlength{\parindent}{0pt}
        \setlength{\parskip}{0pt}
    },
    #1
}

\pgfplotsset{compat=1.18}
\begin{document}
\title{SarcasmReasoner: An LLM Post-Training Framework for Multimodal Sarcasm Reasoning}
% SarcasmReasoner: A Reinforcement Learning Enabled Post-training Framework for Multimodal Sarcasm Reasoning 
% SARCASMREASONER: A Reinforcement Learning based MLLM Post-training Framework for Text-Audio-Visual Sarcasm Reasoning
% SarcasmReasoner: An RL-Based Post-Training Framework for Multimodal Sarcasm Reasoning
% Post-Training Multimodal Large Language Models for Text-Audio-Visual Sarcasm Reasoning
% Multimodal and TAV seem repetitive in terms of information.

\author{
Zhu Li\,\orcidlink{0000-0002-1409-2482},
Yongjian Chen,
Huiyuan Lai\,\orcidlink{0000-0002-5207-6280},
Xiyuan Gao\,\orcidlink{0000-0003-0870-6721},
Shekhar Nayak\,\orcidlink{0000-0002-4277-4851},
and Matt Coler\,\orcidlink{0000-0002-7631-5063}%
\thanks{Zhu Li, Xiyuan Gao, Shekhar Nayak, and Matt Coler are with Campus Fryslân, University of Groningen, 8911 CE Leeuwarden, the Netherlands.}
\thanks{Yongjian Chen is with the Center for Language
and Cognition, University of Groningen, 9718 BG Groningen, the Netherlands.}
\thanks{Huiyuan Lai is with the School of Artificial Intelligence, Nanjing University, Nanjing 210023, China.}
% (e-mail: zhu.li@rug.nl; xiyuan.gao@rug.nl;
% s.nayak@rug.nl; m.coler@rug.nl).}%
\thanks{Manuscript submitted on November 7, 2026.}}

% \thanks{Corresponding author: Zhu Li.}%
% }
% \author{IEEE Publication Technology,~\IEEEmembership{Staff,~IEEE,}
%         % <-this % stops a space
% \thanks{This paper was produced by the IEEE Publication Technology Group. They are in Piscataway, NJ.}% <-this % stops a space
% \thanks{Manuscript received April 19, 2021; revised August 16, 2021.}}

% The paper headers
\markboth{Journal of \LaTeX\ Class Files,~Vol.~14, No.~8, August~2021}%
{Shell \MakeLowercase{\textit{et al.}}: A Sample Article Using IEEEtran.cls for IEEE Journals}

\IEEEpubid{0000--0000/00\$00.00~\copyright~2021 IEEE}
% Remember, if you use this you must call \IEEEpubidadjcol in the second
% column for its text to clear the IEEEpubid mark.
% \thanks{Manuscript submitted on June 13, 2024. \textit{(Corresponding author: Zhu Li)}} 

\maketitle

\begin{abstract}
Multimodal sarcasm detection requires identifying subtle incongruities across textual, acoustic, and visual cues, and remains difficult for current systems. Although multimodal large language models (MLLMs) show potential in modeling emotion and intent, zero-shot evaluations on English (MUStARD++) and Chinese (MCSD) datasets show limitations in sarcasm detection.
To address this issue, we propose SarcasmReasoner, a three-stage reinforcement learning-based post-training framework that reformulates sarcasm detection as evidence-grounded reasoning. In Stage 1, a large omni-modal teacher generates diverse reasoning trajectories through stochastic sampling. In Stage 2, dual-track distillation uses correctly classified, structurally valid trajectories to initialize a smaller student through supervised fine-tuning (SFT), while both correct and incorrect trajectories are scored by modality-specific judges to train a generative reward model (GenRM) for evidence grounding. In Stage 3, Group Relative Policy Optimization (GRPO) combines prediction accuracy, format validity, and GenRM-based grounding rewards.
SarcasmReasoner achieves 72.13\% Macro-F1 on MUStARD++ and 77.09\% on MCSD. These scores are 17.49 and 24.89 percentage points higher than those of its zero-shot Qwen2.5-Omni-7B backbone, respectively. Pairwise human evaluation suggests a preference for GenRM-guided explanations in visual grounding on both datasets. These results suggest that reasoning-aware post-training can improve detection performance while encouraging more evidence-grounded explanations.
\end{abstract}


\section{Introduction}
\IEEEPARstart{S}{arcasm} is a socially and affectively complex form of communication in which a speaker's intended attitude diverges from the literal meaning of an utterance. While sarcasm can be conveyed through linguistic cues, explicit lexical markers are often insufficient \cite{uchiyama2012distinction, castro-etal-2019-towards}. 
Listeners must instead integrate what is said with how it is said and displayed, including prosodic, facial, and gestural cues (e.g., overemphasis on a word or an exaggerated facial expression).
Moreover, recognizing sarcasm frequently depends on detecting contextual incongruity across modalities. For instance, when someone says ``Oh, that’s just great" with a flat intonation and a rolling of the eyes, the literal text alone conveys positivity, but prosody and gestures reveal a sarcastic undertone. Sarcasm understanding therefore requires more than recognizing isolated affective cues: it requires reasoning about whether textual, acoustic, and visual signals jointly support a speaker's implied communicative intent \cite{castro-etal-2019-towards}. 

This cross-modal inference is further complicated by linguistic and cultural variation. The production and interpretation of sarcasm differ across speakers, languages, and communicative communities \cite{joshi2016cultural}. Nevertheless, most multimodal sarcasm detection research has focused on English data and has typically formulated the task as supervised label prediction \cite{ray-etal-2022-multimodal, raghuvanshi2025intra, gao2024amused, li2025evaluating}. 
While previous analyses demonstrate that audio and visual information can complement textual content, many recent state-of-the-art multimodal sarcasm detection approaches primarily evaluate modality contributions through ablation studies, comparing unimodal, bimodal, and trimodal configurations using classification metrics such as accuracy and F1-score. Such performance-oriented comparisons quantify the benefits of different modality combinations but do not explicitly examine how models reason over multimodal evidence or whether their predictions are grounded in observable textual, acoustic, and visual cues. Consequently, improved performance with additional modalities does not necessarily indicate that a model has identified the relevant affective cues or resolved the pragmatic incongruity underlying sarcasm. A model may still provide the correct label through textual shortcuts, dataset-specific correlations, or heuristic assumptions about a speaker's tone or expression \cite{qin-etal-2023-mmsd2,guo-etal-2025-multi,chen2026hear}. Prediction accuracy and modality-level ablations alone are therefore insufficient to assess evidence-grounded sarcasm understanding, motivating approaches that explicitly evaluate the reasoning process and its alignment with observable multimodal evidence.

Recent multimodal large language models (MLLMs) provide an opportunity to move beyond feature fusion toward reasoning over speech, vision, and language \cite{xu2025qwen25omni}. Omni-modal models can process audiovisual input and explain how prosodic, facial, and lexical cues contribute to an interpretation. However, the explanations they generate are not always reliable. An MLLM may provide hallucinated acoustic or visual evidence to justify either a correct or an incorrect prediction \cite{leng2026curse}. For example, claiming that a speaker uses a mocking tone, exaggerated pitch, or an eye roll when no such cue is observable. Such outputs are particularly problematic in affective computing because they may provide seemingly interpretable reasoning without faithful grounding in the human behavior being analyzed.

\IEEEpubidadjcol
To address these challenges, we reformulate multimodal sarcasm detection as an \emph{evidence-grounded affective reasoning} task. We first evaluate recent omni-modal LLMs on two complementary datasets: the English Multimodal Sarcasm Detection Dataset MUStARD++ \cite{ray-etal-2022-multimodal} and the Multimodal Chinese Sarcasm Dataset MCSD \cite{gao2025multimodal}. Results show that zero-shot MLLMs alone are insufficient for multimodal sarcasm detection on these datasets.
Building on this evaluation, we propose \textsc{SarcasmReasoner}, a reinforcement learning (RL)-based post-training framework that combines diverse reasoning trajectory construction, dual-track distillation, and evidence-grounded reward guidance for multimodal sarcasm reasoning. Correctly classified teacher trajectories supervise the student, while both correct and incorrect trajectories train a generative reward model that guides RL toward evidence-grounded explanations.
Post-training enables a 7B omni-modal model to outperform the 30B teacher under zero-shot inference on both datasets. 
Ablation studies show that Diverse-8 provides the strongest SFT initialization among the evaluated strategies, while GenRM-based reward guidance mainly affects evidence grounding.
Pairwise human evaluation suggests improvements in modality-specific evidence grounding on both datasets, particularly for visual evidence, while qualitative error analysis illustrates how unsupported multimodal interpretations can affect predictions.

The main contributions of this work are as follows:

\begin{itemize}
    \item We propose \textsc{SarcasmReasoner}, a three-stage post-training framework that constructs diverse teacher-generated reasoning trajectories, performs supervised reasoning distillation, and applies multi-objective reward-guided policy optimization for multimodal sarcasm understanding.

    \item We use stochastic teacher sampling to construct a diverse set of multimodal reasoning trajectories instead of relying on a single deterministic teacher path, exposing the student model to multiple reasoning patterns for relating textual, acoustic, and visual cues to sarcastic intent.

    \item We introduce a dual-track distillation strategy that uses correctly classified trajectories to supervise the student model while retaining both correct and incorrect trajectories to train a generative reward model for modality-specific evidence grounding.

    \item We find that Diverse sampling achieves the strongest SFT performance among the evaluated trajectory construction strategies, while pairwise human evaluation suggests that GenRM-guided reward alignment improves evidence grounding, particularly for visual evidence.
\end{itemize}

Overall, this study extends RL-based post-training to multimodal pragmatic inference, where reliable explanations require grounding in subtle and potentially conflicting signals across text, speech, and vision.

\section{Related Work}

\subsection{Multimodal Sarcasm Detection}

Early research on multimodal sarcasm detection primarily focused on \textit{visual-textual sarcasm detection}, where images and captions jointly convey ironic meaning \cite{schifanella2016detecting, qin-etal-2023-mmsd2, zhang2025sarcasmbench}. These studies typically encoded each modality separately and then fused the resulting higher-level features, using methods that ranged from simple concatenation \cite{schifanella2016detecting} to attention-based modeling of inter- and intra-modal incongruities \cite{pan2020modeling}. More recent approaches have adopted multimodal encoders based on joint image-text representations or contrastive image-text learning \cite{li2019visualbert, radford2021learning}, or integrated large language models through prompt engineering for sarcasm detection \cite{ding2022multi}. Despite these advances, much of this work remains centered on text-only or visual-textual scenarios, leaving sarcasm in natural spoken interactions comparatively underexplored.

Sarcasm in videos, such as sitcoms and stand-up comedy, requires reasoning across speech, facial expressions, gestures, and textual transcripts. Prior work on datasets such as MUStARD \cite{castro-etal-2019-towards} and MUStARD++ \cite{ray-etal-2022-multimodal} has shown that audio and visual cues can significantly enhance sarcasm detection. Subsequent methods have leveraged multimodal attention \cite{wu2021modeling, aggarwal2023multimodal}, optimal-transport alignment \cite{pramanick2022multimodal}, and collaborative gating fusion \cite{ray-etal-2022-multimodal, raghuvanshi2025intra} to improve performance. Other recent systems further demonstrate the effectiveness of task-specific architectures for multimodal sarcasm detection \cite{gao2024amused}. However, these approaches primarily rely on supervised classification objectives and task-specific fusion architectures rather than explicitly generating and evaluating structured cross-modal reasoning. They therefore provide limited insight into how omni-modal foundation models can be adapted to produce evidence-grounded explanations for spoken sarcasm.

\begin{figure*}[t]
    \centering
    \includegraphics[width=\linewidth]{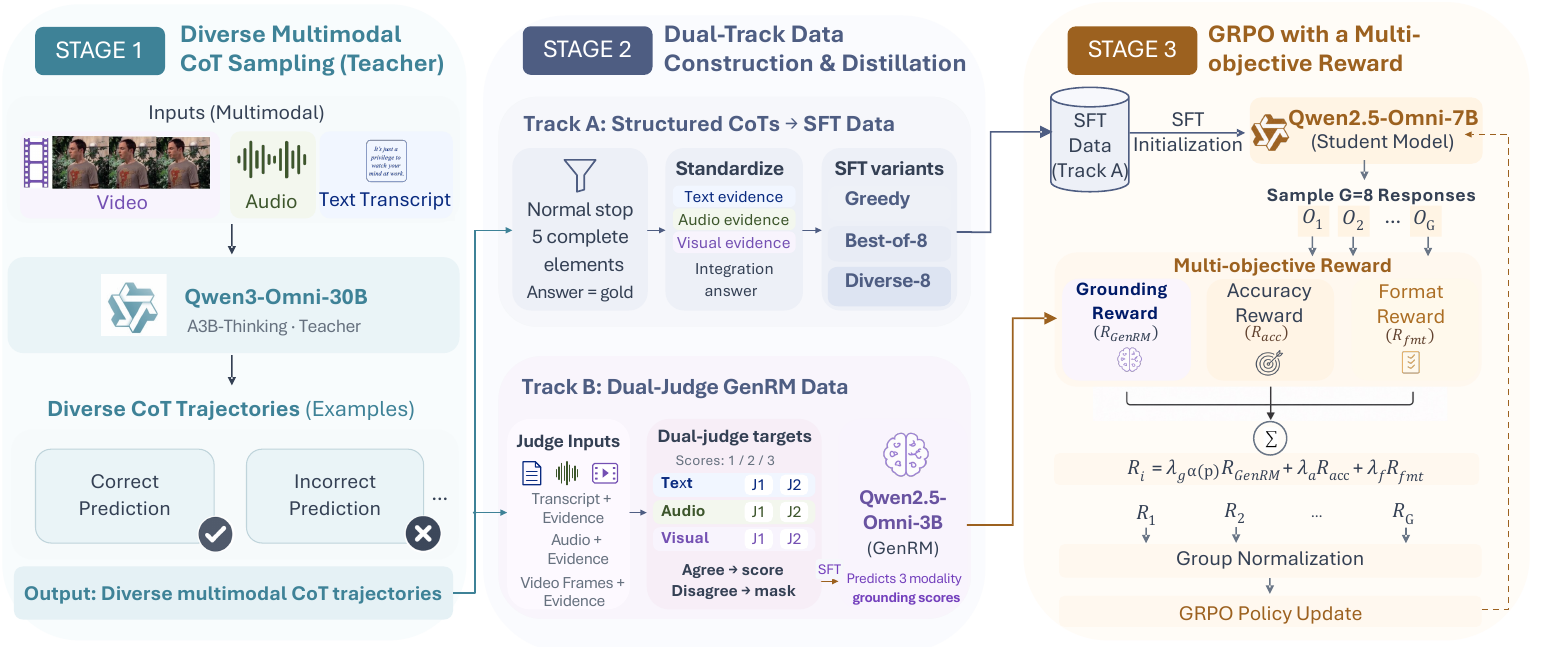}
    \caption{Overview of the proposed SarcasmReasoner framework.}
    \label{fig:arch}
\end{figure*}

\subsection{MLLMs for Sarcasm Understanding}

General-purpose MLLMs have recently demonstrated strong capabilities across text, audio, and visual understanding tasks. Large-scale speech and multimodal foundation models have shown strong performance in speech recognition, synthesis, and cross-modal understanding \cite{radford2023robust, hurst2024gpt, chu2023qwen}. More recent omni-modal models further suggest that a single foundation model can integrate information across speech, vision, and text \cite{xu2025qwen25omni}. These multimodal understanding capabilities open new opportunities for tasks involving complex cross-modal incongruities.

However, it remains unclear whether the multimodal reasoning abilities of MLLMs extend to fine-grained pragmatic phenomena such as conversational sarcasm. Existing MLLM benchmarks primarily evaluate meme-based or broader vision-language content understanding rather than sarcasm reasoning in spoken interactions \cite{lin2024goat, yang2023mm}. As a result, they leave open the question of how well MLLMs handle sarcasm in spoken conversational contexts, where pragmatic meaning often depends on subtle interactions among lexical content, prosody, facial expressions, and gestures.

\subsection{Post-training for Grounded Multimodal Reasoning}

Although pre-trained foundation models provide strong multimodal representations, pre-training alone does not guarantee reliable reasoning over complex pragmatic phenomena in conversational settings. Post-training techniques such as SFT and RL have therefore been widely used to adapt foundation models to downstream tasks requiring structured reasoning \cite{schulman2017proximal, ouyang2022training, shao2024deepseekmath}. While SFT can improve task performance and align models with desired response formats, prior work on the superficial alignment hypothesis suggests that such alignment may primarily teach models how to express existing capabilities in the expected format, rather than ensuring faithful task-specific reasoning \cite{zhou2023lima}. Complementarily, post-training methods such as Reinforcement Learning from Human Feedback (RLHF) and reward-guided RL have been widely explored to further improve instruction following and reasoning-oriented behavior in LLMs \cite{ouyang2022training, guo2025deepseek}.

Multimodal sarcasm detection is a particularly challenging testbed for such post-training. Unlike conventional emotion recognition, sarcasm often involves pragmatic incongruity between literal meaning and non-verbal signals such as prosody, facial expression, and contextual intent. For example, positive lexical content delivered with exaggerated or flat prosody may indicate sarcastic intent \cite{castro-etal-2019-towards, ray-etal-2022-multimodal}. Interpreting such cues motivates structured multimodal reasoning, often elicited through chain-of-thought (CoT) prompting \cite{wei2022chain, kojima2022large}. However, generating CoT explanations does not guarantee that the described multimodal evidence is supported by the input, as LLMs may still produce hallucinated claims \cite{huang2025survey}. This highlights the need for reasoning-aware post-training methods that not only improve prediction accuracy but also encourage evidence-grounded explanations.

\subsection{Reasoning-aware Sarcasm and Speech Understanding}

Recent work has begun to incorporate reasoning supervision into sarcasm detection. One approach enhances textual sarcasm detection through focused reasoning strategies, showing that structured reasoning can improve interpretability and robustness \cite{yang2025sarcasm}. However, its formulation is primarily text-centric and does not explicitly address multimodal grounding. Another study introduced a diagnostic benchmark for evaluating pragmatic reasoning in video language models, disentangling perception and reasoning errors and proposing a pragmatic reasoning framework to steer models toward intention-focused inference \cite{saha2025mustreason}. While this work provides useful analysis of perception-reasoning gaps, it does not explore RL-based post-training or mechanisms for mitigating hallucinated multimodal evidence.

Parallel efforts in speech affective understanding have also begun to explore reasoning-centric RL strategies. One approach reformulates speech emotion recognition as a prosody-aware reasoning task and combines prosody-enhanced modeling with reward-guided policy optimization to improve interpretability and reasoning quality \cite{wang2026emotionthinker}. 
Related approaches, such as GRPO-guided modality selection \cite{chen2025grpo} and B-GRPO \cite{gao2026b}, further demonstrate the potential of policy optimization for adaptive modality usage and unsupervised sample selection. These studies suggest that RL-based post-training can improve reasoning-oriented outputs beyond simple label prediction.

Nevertheless, extending these strategies to multimodal sarcasm detection is non-trivial. First, sarcasm often relies on subtle cross-modal incongruity rather than explicit affective cues. Second, MLLMs may generate hallucinated acoustic or visual evidence even when their final predictions are correct. Third, while reasoning-annotated datasets have been developed for speech emotion recognition \cite{wang2026emotionthinker}, comparable large-scale resources for multimodal sarcasm reasoning remain limited, limiting direct reasoning supervision. 
These challenges motivate our post-training framework, presented in the following section, which combines teacher-generated reasoning supervision with reward guidance for modality-specific evidence grounding.

% \section{Method}
\section{The SarcasmReasoner Framework} % SarcasmReasoner: A reinforcement -learning based post training framework for multimodal sarcasm reasoning
We propose SarcasmReasoner, a three-stage post-training framework for evidence-grounded multimodal sarcasm reasoning (Figure~\ref{fig:arch}). Stage~1 samples diverse reasoning trajectories from a multimodal teacher. Stage~2 constructs two complementary training sets: correctly classified trajectories are distilled into the student through SFT, while complete trajectories are scored by two independent judges per modality to provide grounding supervision for a multimodal generative reward model (GenRM). In Stage~3, we further optimize the student with GRPO, combining prediction correctness and format rewards with an additional GenRM-based grounding reward.

\subsection{Stage 1: Diverse Multimodal CoT Sampling}
Multimodal sarcasm datasets typically lack multi-step reasoning annotations. To obtain reasoning supervision, we prompt a powerful teacher model $\mathcal{M}_{\mathrm{teacher}}$ (Qwen3-Omni-30B-A3B-Thinking) to analyze textual, acoustic, and visual evidence separately before integrating the available cues to predict sarcasm. Given a multimodal input $x=(v,a,t)$, where $v$, $a$, and $t$ denote video frames, audio signals, and text transcripts, respectively, the teacher generates a structured reasoning trajectory containing five elements: text evidence, audio evidence, visual evidence, integration, and a final sarcasm prediction. 
% The prompt encourages evidence-grounded reasoning without assuming that every modality necessarily provides cues of sarcasm. Any additional free-form teacher thinking is excluded from the student training targets.
Instead of relying on a single deterministic trajectory, we independently sample $N=8$ reasoning trajectories per input from the teacher's conditional distribution. This produces a diverse pool of reasoning paths containing both correct and incorrect predictions, which are subsequently used for dual-track distillation.
The complete output format and sampling parameters are provided in Appendix~\ref{app:sft_data}.

% Instead of generating a single deterministic trajectory via greedy decoding, we sample a diverse pool of reasoning paths from the teacher's conditional distribution for each input $x$. Specifically, we independently sample $N=8$ trajectories as $y_i \sim P_{\mathcal{M}_{\text{teacher}}}(\cdot|x)$, using temperature-based sampling (temperature $T=0.6$, $\text{top\_p}=0.95$), forming a trajectory set $\mathcal{Y}=\{y_1,\dots,y_N\}$. This stochastic process produces a heterogeneous reasoning pool containing both correct and incorrect predictions, providing diverse supervision for subsequent dual-track distillation.

\subsection{Stage 2: Dual-track Data Construction \& Distillation}
Rejection-based fine-tuning typically retains correctly classified trajectories and discards incorrect ones, potentially overlooking useful supervision in the discarded responses. To address this limitation, we propose a dual-track distillation strategy that uses correctly classified trajectories for student initialization while retaining both correct and incorrect trajectories for grounding assessment.

\noindent\textbf{Track A: Filtered Correct Trajectories for SFT.}
To construct supervised reasoning data, we retain teacher-generated trajectories that have correct final predictions and complete structured outputs.
We then compare three SFT data construction strategies that differ in how teacher responses are generated and selected. \emph{Greedy} uses deterministic decoding to produce a single trajectory per input. In contrast, \emph{Best-of-8} and \emph{Diverse-8} both use stochastic decoding to independently sample eight candidate trajectories per input. % with temperature $T=0.6$ and top-$p=0.95$. 
Best-of-8 selects the eligible trajectory with the highest mean token log-probability, retaining only one response per input. Diverse-8 instead retains all eligible trajectories from the same sampled pool, allowing multiple reasoning paths to provide supervision for the same input. 
% For all strategies, inputs without any eligible trajectories are excluded from the corresponding SFT dataset.
These strategies provide different forms of supervision. Compared with Greedy, Best-of-8 can increase the coverage of training inputs by selecting from multiple candidates, while Diverse-8 additionally exposes the student to multiple valid reasoning trajectories for the same input. 
Their resulting dataset sizes and coverage are reported in Appendix~\ref{app:sft_data}.
In the main pipeline, we adopt Diverse-8 to construct $\mathcal{D}_{\mathrm{SFT}}$ and fine-tune Qwen2.5-Omni-7B, obtaining the initial student policy $\pi_{\mathrm{SFT}}$ for subsequent reinforcement learning.

\noindent\textbf{Track B: Generative Reward Model Training.}
A correct final prediction does not necessarily guarantee that the intermediate reasoning is grounded in observable multimodal evidence. We therefore train a generative reward model (GenRM) to assess the grounding quality of modality-specific evidence independently of prediction correctness. Specifically, we construct a grounding-judge dataset $\mathcal{D}_{\mathrm{Judge}}$ using complete teacher-generated trajectories with both correct and incorrect final predictions. Each trajectory is evaluated by two independent judges per modality, who assess whether the generated textual, acoustic, and visual evidence is supported by the corresponding input. Each judge assigns a score of 1 (unreliable), 2 (partially reliable), or 3 (reliable). For each modality, a score is retained as supervision only when both judges agree; otherwise, the corresponding supervision is masked. Trajectories with at least one agreed modality score are retained, allowing the GenRM to learn from partially available grounding supervision.
We fine-tune Qwen2.5-Omni-3B on $\mathcal{D}_{\mathrm{Judge}}$ to obtain GenRM, which predicts the three modality-specific grounding scores from the original multimodal input and generated evidence. The gold label, final prediction, and integration reasoning are excluded to focus the assessment on evidence grounding. During training, only agreed scores contribute to the loss. At inference time, the predicted scores are mapped from $\{1,2,3\}$ to $\{0,0.5,1\}$ and averaged to provide a grounding reward.  The judge models, agreement criterion, and supervision details are provided in Appendix~\ref{app:judge_data}.

\subsection{Stage 3: GRPO with a Multi-objective Reward}

Building upon the SFT initialization, we optimize the student model using GRPO. Answer-level optimization alone does not constrain how a prediction is justified: a model may predict the correct label while generating unsupported acoustic or visual evidence. To address this issue, we compare two GRPO variants initialized from the same SFT checkpoint. Standard GRPO uses accuracy and format rewards, while GenRM-guided GRPO additionally incorporates a progressively weighted grounding reward.

For each multimodal input $x$, the policy samples $G=8$ reasoning trajectories. The total reward for each trajectory $o_i$ at training progress $p$ is defined as:
\begin{equation}
\begin{aligned}
R_i(p) &= \lambda_a R_{\mathrm{acc}}(o_i)
       + \lambda_f R_{\mathrm{fmt}}(o_i) \\
       &\quad + \lambda_g \alpha(p) R_{\mathrm{GenRM}}(x,o_i),
\end{aligned}
\end{equation}
where $R_{\mathrm{acc}}$ rewards correct predictions, $R_{\mathrm{fmt}}$ rewards valid structured outputs, and $R_{\mathrm{GenRM}}$ evaluates evidence grounding. The coefficients $\lambda_a$, $\lambda_f$, and $\lambda_g$ control their relative contributions, while $\alpha(p)$ gradually increases from 0 to 1 during GenRM-guided optimization, progressively introducing the grounding reward.
The frozen GenRM predicts a grounding score of 1, 2, or 3 for each modality. We normalize these scores to $\{0,0.5,1\}$ and average them to obtain the grounding reward. This reward is applied only when the final prediction is correct and all three evidence sections are complete; otherwise, it is set to zero. 
This correctness-based gating prevents high grounding scores from compensating for incorrect predictions, although it also introduces an additional accuracy-dependent reward signal beyond the standard GRPO reward.
During GenRM-guided optimization, the grounding reward initially has zero weight, then increases linearly to its full weight and remains constant thereafter according to $\alpha(p)$. GRPO uses group-relative rewards to update the student policy, with KL regularization constraining policy drift. Detailed reward definitions, scheduling, and training parameters are provided in Appendix~\ref{app:training_setup}.

\section{Experimental Setup}
\noindent\textbf{Dataset.} 
We evaluate the proposed SarcasmReasoner framework on two datasets: 1) MUStARD++ \cite{ray-etal-2022-multimodal}, an English multimodal sarcasm detection dataset containing 1,202 labeled utterances with text, speech, and video modalities, including 601 sarcastic and 601 non-sarcastic samples.
2) \textbf{MCSD} \cite{gao2025multimodal}, a Chinese Multimodal Sarcasm Dataset collected from Chinese stand-up comedy. It contains 2,705 aligned video, audio, and manually transcribed utterances, including 1,344 sarcastic and 1,361 non-sarcastic samples.
We partition both datasets into training, validation, and test sets with an approximately 70/15/15 ratio. 
Detailed dataset statistics are provided in Appendix~\ref{app:datasets}.

\noindent\textbf{Grounding Judges.}
To construct the grounding supervision for GenRM, we employ two independent judge models for each modality: Gemma-3-12B-IT \cite{gemmateam2025gemma3} and DeepSeek-V2-Lite-Chat \cite{deepseekai2024deepseekv2} for text, Step-Audio-2-mini \cite{wu2025stepaudio2} and Phi-4-Multimodal-Instruct \cite{abouelenin2025phi} for audio, and InternVL3.5-8B \cite{wang2025internvl35} and Molmo2-8B \cite{clark2026molmo2} for visual evidence. Each judge evaluates whether the observations and interpretations in the generated evidence are supported by the corresponding input modality, assigning a score of 1 (unreliable), 2 (partially reliable), or 3 (reliable). Only scores agreed upon by both judges are used for GenRM supervision; disagreements are masked. 
The shared judging rubric and score distributions are provided in Appendix~\ref{app:judge_data}.

\noindent\textbf{Evaluation.}
We evaluate the proposed framework in terms of sarcasm classification performance and evidence grounding. For classification, we report macro-averaged F1-score (Macro-F1) and class-specific F1 scores for sarcastic and non-sarcastic samples.
We compare Diverse-8 SFT, SFT+GRPO, and SarcasmReasoner using two-sided exact McNemar tests \cite{mcnemar1947note}, with Holm correction across six comparisons, and estimate pointwise 95\% confidence intervals for Macro-F1 differences using 100,000 paired bootstrap resamples.
To evaluate GenRM, we first measure its ability to predict the three-level grounding scores derived from dual-judge agreement on the validation sets. We report Macro-F1 separately for text, audio, and visual evidence, together with their unweighted average.
We further use the trained GenRM to assess evidence grounding in the outputs generated by three representative post-training systems: SFT, SFT+GRPO, and the proposed SarcasmReasoner (SFT+GRPO+GenRM). These scores provide a model-based assessment of evidence grounding.
For human evaluation, we selected 50 samples from each dataset using label-stratified sampling. Annotators were shown only the modality-specific evidence sections from two anonymized model outputs. 
For each system pair, textual, acoustic, and visual evidence is assessed separately. Annotators select the explanation better supported by the corresponding input modality or indicate a tie. System identities are concealed, and presentation order is randomized.  Annotators are not shown the gold sarcasm labels. All annotators are native Mandarin Chinese speakers with high proficiency in English. Before the human evaluation, all three annotators were informed of the study purpose and evaluation procedures and provided informed consent to participate.
We aggregate the three annotators' individual judgments, yielding up to 150 preferences per model pair and modality. 
Inter-annotator agreement on pairwise preferences for evidence grounding in reasoning outputs is assessed using Fleiss' $\kappa$, yielding overall values of 0.647 on MUStARD++ and 0.655 on MCSD. 
The modality-specific human evaluation interfaces and modality-specific agreement scores are reported in Appendix~\ref{app:human_protocol}.
% not a kappa score for usual sarcastic annotation, but for the reasoning output ratings/scores

\noindent\textbf{Cross-lingual Training Settings.}
Since our study covers English and Chinese sarcasm, we additionally investigate cross-lingual transfer and the effects of language balance in training. We first consider five SFT settings: English-only, Chinese-only, pooled English+Chinese, language-balanced downsampling, and language-balanced oversampling. The monolingual settings allow us to examine cross-lingual transfer by evaluating each model on both languages. For bilingual SFT, the pooled setting combines both datasets directly, while downsampling and oversampling balance their sample counts by reducing Chinese or increasing English training examples, respectively.
We further investigate whether RL improves bilingual sarcasm detection by applying GRPO and GenRM-guided GRPO to the pooled bilingual SFT model. All models are evaluated on both the English and Chinese test sets. We report Macro-F1 separately for each language, together with their unweighted average.

\noindent\textbf{Baselines and Model Variants.} We compare SarcasmReasoner with task-specific supervised baselines and zero-shot omni-modal language models. For task-specific supervised comparison, we include an SVM-based classifier \cite{castro-etal-2019-towards} and the collaborative gating model \cite{ray-etal-2022-multimodal}, following prior work on multimodal sarcasm detection\footnote{Both baselines are retrained on our training splits and evaluated on the same test splits as SarcasmReasoner.}. For zero-shot evaluation, we consider Phi-4-Multimodal \cite{abouelenin2025phi}, MiniCPM-o-2.6 \cite{yao2024minicpm}, Qwen2.5-Omni-7B \cite{xu2025qwen25omni}, Qwen3-Omni-30B-A3B-Instruct, and Qwen3-Omni-30B-A3B-Thinking \cite{Qwen3-Omni}.
We further conduct stage-wise comparisons among different post-training variants of SarcasmReasoner. For SFT, we compare students trained on trajectories constructed using Greedy, Best-of-8, and Diverse-8 strategies. For RL-based post-training, we evaluate GRPO without SFT initialization, GRPO initialized from each SFT variant and trained using accuracy and format rewards, and the full SarcasmReasoner configuration with GenRM-guided optimization.

\begin{table*}[t]
\centering
\small
\setlength{\tabcolsep}{4pt}
\renewcommand{\arraystretch}{1.12}
\caption{Main sarcasm detection results on MUStARD++ and MCSD.
Sarc-F1 and Non-sarc-F1 denote F1 scores for the sarcastic
and non-sarcastic classes, respectively.
All scores are reported as percentages (\%).
Bold indicates the best result in each column.}
\label{tab:main_results_combined}

\begin{tabular*}{\textwidth}
{@{\extracolsep{\fill}}lcccccc@{}}
\toprule
\multirow{2}{*}{\textbf{Model}}
& \multicolumn{3}{c}{\textbf{MUStARD++}}
& \multicolumn{3}{c}{\textbf{MCSD}} \\
\cmidrule(lr){2-4} \cmidrule(lr){5-7}
& \textbf{Macro-F1}
& \textbf{Sarc-F1}
& \textbf{Non-sarc-F1}
& \textbf{Macro-F1}
& \textbf{Sarc-F1}
& \textbf{Non-sarc-F1} \\
\midrule

\multicolumn{7}{l}{\textit{Task-specific supervised models}} \\
SVM
& 70.14 & 70.97 & 69.32
& 73.66 & 71.20 & 76.13 \\
Collaborative gating
& 67.74 & 72.73 & 62.75
& 73.68 & 71.35 & 76.02 \\
\midrule

\multicolumn{7}{l}{\textit{Omni-modal LLMs (zero-shot)}} \\
MiniCPM-o-2.6
& 57.55 & 64.76 & 50.33
& 59.31 & 60.39 & 58.23 \\
% MiniCPM-o-4.5
% & 64.85 & 61.35 & 68.34
% & 57.88 & 57.99 & 57.78 \\
Phi-4-Multimodal
& 50.13 & 34.15 & 66.11
& 57.98 & 48.23 & 67.74 \\
Qwen2.5-Omni-7B
& 54.64 & 61.90 & 47.37
& 52.20 & 62.63 & 41.77 \\
% Qwen3-Omni-30B Instruct
% & 60.77 & 61.20 & 60.34
% & 61.24 & 57.61 & 64.86 \\
Qwen3-Omni-30B Thinking
& 63.13 & 67.00 & 59.26
& 64.53 & 64.88 & 64.18 \\
\midrule

\multicolumn{7}{l}{
\textit{Post-training (Qwen2.5-Omni-7B backbone)}} \\
\textbf{SarcasmReasoner (Ours)}
& \textbf{72.13} & \textbf{74.75} & \textbf{69.51}
& \textbf{77.09} & \textbf{76.81} & \textbf{77.37} \\

\bottomrule
\end{tabular*}
\end{table*}

\noindent\textbf{Implementation Details.}
We use the \texttt{ms-swift} fine-tuning framework \cite{zhao2024swiftascalablelightweightinfrastructure} with the AdamW optimizer and LoRA adapters ($r=8$, $\alpha_{\mathrm{LoRA}}=32$) \cite{hu2022lora}.
All SFT variants use a learning rate of $10^{-4}$, with training duration determined by validation-loss convergence for each strategy.
The Qwen2.5-Omni-3B GenRM undergoes three epochs of SFT with a learning rate of $2\times10^{-5}$.
Both GRPO and SarcasmReasoner start from the SFT-trained student and run for two epochs with $G=8$ trajectories per query, a learning rate of $10^{-5}$, and a KL coefficient of $\beta=0.04$.
GRPO uses accuracy and format rewards, while SarcasmReasoner additionally includes the GenRM grounding reward.
We set $\lambda_a=1$ and $\lambda_f=\lambda_g=0.1$,
with the grounding reward further scaled by $\alpha(p)$, where $p\in[0,1]$ denotes SarcasmReasoner training progress.
The multiplier $\alpha(p)$ is zero for $p\leq0.5$,
increases linearly to one over $0.5<p<0.6$,
and remains one thereafter.
Each run uses a single NVIDIA H100 GPU. %  using seed 42
Detailed training hyperparameters are provided in Table~\ref{tab:training_details} in Appendix~\ref{app:training_setup}.

\section{Results}
\subsection{Main Results: Supervised, Zero-Shot, and Post-Training Comparisons}

Table~\ref{tab:main_results_combined} summarizes the main detection results on MUStARD++ and MCSD.
The conventional supervised baselines remain competitive on both datasets.
The SVM achieves 70.14\% Macro-F1 on MUStARD++, while collaborative gating achieves 73.68\% on MCSD.
Among the zero-shot omni-modal LLMs, MiniCPM-o-2.6 achieves 57.55\% Macro-F1 on MUStARD++ and 59.31\% on MCSD, while Phi-4-Multimodal reaches 50.13\% and 57.98\%, respectively. Qwen2.5-Omni-7B obtains 54.64\% on MUStARD++ and 52.20\% on MCSD. The strongest zero-shot performance is achieved by Qwen3-Omni-30B Thinking, with 63.13\% Macro-F1 on MUStARD++ and 64.53\% on MCSD.
Despite these improvements, all zero-shot omni-modal LLMs remain below the strongest task-specific supervised baselines, particularly on MCSD, indicating that general-purpose multimodal capabilities alone are insufficient for robust sarcasm detection.

Task-specific post-training substantially improves the Qwen2.5-Omni-7B backbone.
The full SarcasmReasoner configuration, SFT+GRPO+GenRM, achieves 72.13\% Macro-F1 on MUStARD++ and 77.09\% on MCSD.
On MUStARD++, the observed Macro-F1 is 1.99 percentage points higher than that of the strongest conventional supervised baseline (70.14\%). On MCSD, the corresponding difference is 3.41 percentage points.
Notably, the post-trained 7B model also outperforms the 30B zero-shot model on both datasets, indicating that post-training can offset, and in this setting even surpass, the advantage of increased model scale under zero-shot inference.
Overall, these results show that task-specific post-training can substantially improve multimodal sarcasm detection beyond zero-shot inference while achieving numerically higher Macro-F1 scores than the strongest conventional supervised approaches on both datasets.
We next examine how individual stages of the proposed post-training pipeline contribute to these improvements.

\begin{table*}[t]
\centering
\small
\setlength{\tabcolsep}{4pt}
\renewcommand{\arraystretch}{1.12}
\caption{Stage-wise results on MUStARD++ and MCSD. Stage~1 compares zero-shot teacher variants,
Stage~2 compares trajectory selection strategies for SFT, and Stage~3 compares GRPO with different SFT initializations, with GenRM added to the Diverse-8 configuration. Sarc-F1 and Non-sarc-F1 denote F1 scores for the sarcastic and non-sarcastic classes, respectively. All scores are reported as percentages (\%). Bold indicates the best result in each column.}
\label{tab:stagewise_ablation}

\begin{tabular*}{\textwidth}
{@{\extracolsep{\fill}}llcccccc@{}}
\toprule
\multirow{2}{*}{\textbf{Stage}}
& \multirow{2}{*}{\textbf{Configuration}}
& \multicolumn{3}{c}{\textbf{MUStARD++}}
& \multicolumn{3}{c}{\textbf{MCSD}} \\
\cmidrule(lr){3-5} \cmidrule(lr){6-8}
& & \textbf{Macro-F1}
  & \textbf{Sarc-F1}
  & \textbf{Non-sarc-F1}
  & \textbf{Macro-F1}
  & \textbf{Sarc-F1}
  & \textbf{Non-sarc-F1} \\
\midrule

\multirow{2}{*}{Stage~1}
& Qwen3-Omni-30B Instruct
& 60.77 & 61.20 & 60.34
& 61.24 & 57.61 & 64.86 \\
& Qwen3-Omni-30B Thinking
& 63.13 & 67.00 & 59.26
& 64.53 & 64.88 & 64.18 \\
\midrule

\multirow{3}{*}{Stage~2}
& SFT (Greedy)
& 68.50 & 68.16 & 68.85
& 66.53 & 62.57 & 70.48 \\
& SFT (Best-of-8)
& 63.64 & 59.63 & 67.66
& 68.06 & 65.78 & 70.34 \\
& SFT (Diverse-8)
& 69.05 & 69.57 & 68.54
& 69.82 & 64.69 & 74.95 \\
\midrule

\multirow{5}{*}{Stage~3}
& GRPO (w/o SFT)
& 58.53 & 69.30 & 47.76
& {64.92} & {70.36} & {59.48} \\
& GRPO (Greedy SFT)
& 67.84 & 69.79 & 65.88
& 69.06 & 65.56 & 72.57 \\
& GRPO (Best-of-8 SFT)
& 68.90 & 66.67 & \textbf{71.13}
& 70.94 & 68.29 & 73.59 \\
& GRPO (Diverse-8 SFT)
& 70.16 & 70.33 & 70.00
& 75.93 & 73.85 & \textbf{78.00} \\
& \textbf{SarcasmReasoner (Ours)}
& \textbf{72.13} & \textbf{74.75} & 69.51
& \textbf{77.09} & \textbf{76.81} & 77.37 \\

\bottomrule
\end{tabular*}
\end{table*}

\subsection{Stage-wise Ablation}

\begin{figure}[t]
\centering
\includegraphics[width=0.99\columnwidth]{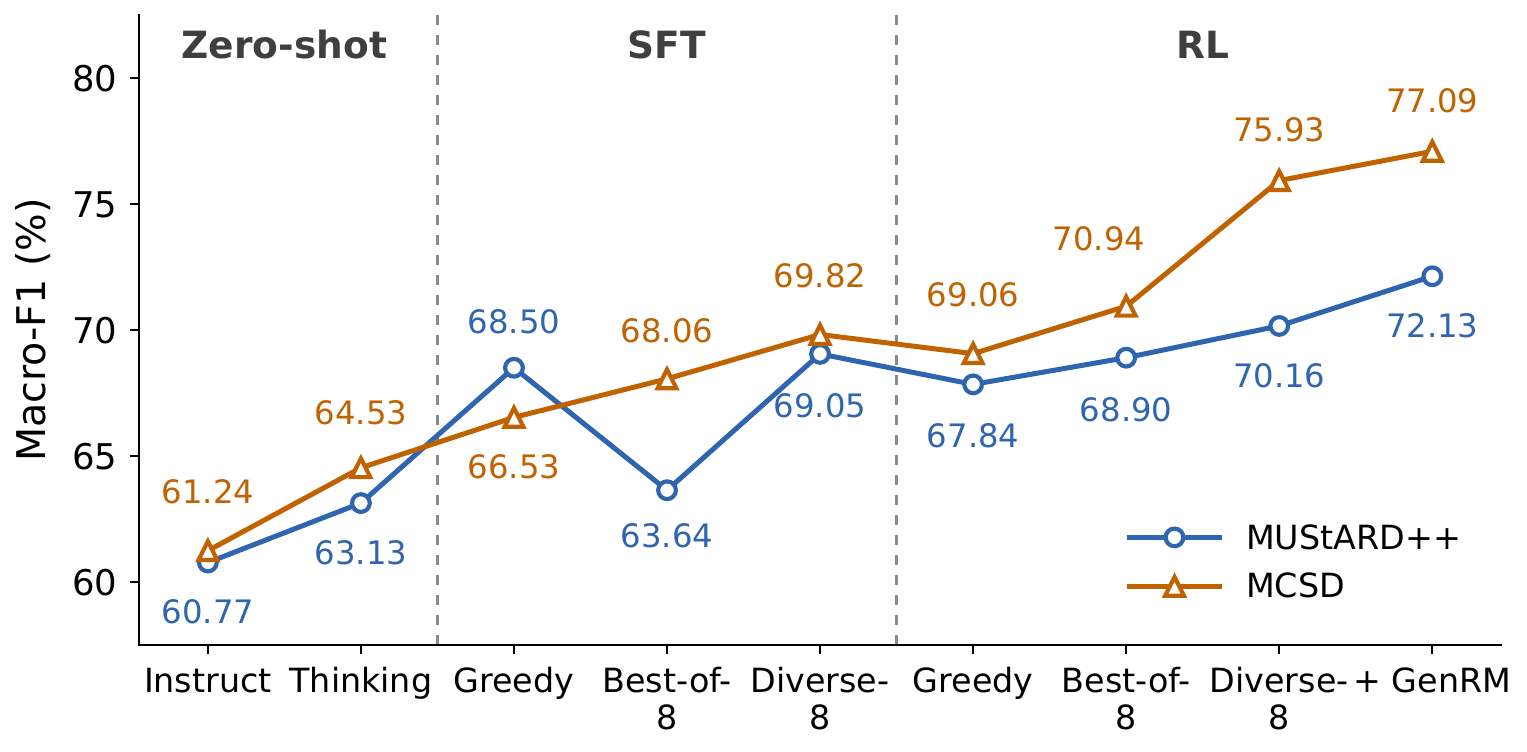}
\caption{Stage-wise Macro-F1 improvements from zero-shot teacher evaluation to supervised reasoning distillation and GRPO-based reward alignment. Dashed vertical lines separate the three phases. MCSD shows larger overall improvements, particularly during RL-based post-training, although the magnitude of improvement may also reflect dataset-specific characteristics.}
\label{fig:stagewise_improvement}
\end{figure}

Table~\ref{tab:stagewise_ablation} and Figure~\ref{fig:stagewise_improvement} analyze the stage-wise performance gains and highlight three consistent patterns across both datasets.
We examine three components: teacher reasoning mode, SFT trajectory construction, and GRPO-based reward alignment.

\noindent\textbf{Effect of Teacher Reasoning Mode.}
We first compare the standard instruction mode and thinking mode of Qwen3-Omni-30B.
Thinking mode consistently improves detection performance on both datasets.
On MUStARD++, Macro-F1 increases from 60.77\% to 63.13\%.
On MCSD, Macro-F1 increases from 61.24\% to 64.53\%.
These results indicate that thinking mode achieves stronger zero-shot performance on both datasets, potentially by producing more structured and informative reasoning trajectories. We therefore adopt it for subsequent trajectory construction.

\noindent\textbf{Effect of SFT Trajectory Construction.}
We next compare three strategies for constructing the SFT initialization data: Greedy, Best-of-8, and Diverse-8.
Best-of-8 improves over greedy decoding on MCSD but performs worse on MUStARD++, suggesting that selecting a single high-probability trajectory does not consistently benefit supervised distillation.
The strongest SFT initialization is obtained with diverse sampling, achieving 69.05\% Macro-F1 on MUStARD++ and 69.82\% on MCSD. These results suggest that retaining multiple valid reasoning trajectories per input can provide effective supervision for multimodal sarcasm detection.
% This suggests that exposure to a larger and more varied set of structured multimodal explanations is beneficial, potentially because it increases reasoning diversity. % beneficial for learning of multimodal sarcasm cues.
% By retaining multiple valid reasoning paths for the same input, diverse sampling exposes the student to a broader range of multimodal reasoning patterns, which may help it learn more flexible associations between sarcasm and textual, acoustic, and visual evidence.

\noindent\textbf{GenRM Validation.}
We evaluate the ability of GenRM to predict modality-specific grounding scores provided by the dual-judge supervision. Validation is performed only on modality scores for which both judges agree. As shown in Table~\ref{tab:genrm_valid_main}, GenRM achieves average validation Macro-F1 scores of 82.82\% on MUStARD++ and 79.25\% on MCSD. Audio grounding scores are predicted most accurately on both datasets (90.18\% and 94.55\%), while text grounding score prediction is more challenging on MCSD (62.68\%). These results indicate that GenRM can learn to approximate the judge-derived grounding scores, providing a basis for its use as an auxiliary reward during GRPO.
%%% future work: check the judges against human ratings (even on just a small sample)
% Validation is also performed only on agreed modality scores, which limits the scope of the validation results.

\begin{table}[t]
\centering
\caption{Validation Macro-F1 (\%) of GenRM.
The average is computed across text, audio, and visual components.}
\label{tab:genrm_valid_main}
\begin{tabular}{lrrrr}
\toprule
\textbf{Dataset} &
\textbf{Text} &
\textbf{Audio} &
\textbf{Visual} &
\textbf{Avg.} \\
\midrule
MUStARD++ & 76.85 & 90.18 & 81.42 & \textbf{82.82} \\
MCSD      & 62.68 & 94.55 & 80.53 & \textbf{79.25} \\
\bottomrule
\end{tabular}
\end{table}

\noindent\textbf{Effect of GRPO-based Alignment.}
Finally, we examine the contribution of RL-based post-training. Applying GRPO directly without SFT initialization (GRPO w/o SFT) achieves only 58.53\% Macro-F1 on MUStARD++ and 64.92\% on MCSD, substantially below GRPO initialized from Diverse-8 SFT (70.16\% and 75.93\%, respectively). These results highlight the importance of supervised reasoning initialization for subsequent policy optimization. % than applying GRPO directly to the zero-shot policy

Starting from the Diverse-8 SFT model, standard GRPO improves Macro-F1 from 69.05\% to 70.16\% on MUStARD++ and from 69.82\% to 75.93\% on MCSD, while SarcasmReasoner reaches 72.13\% and 77.09\%, respectively. On MCSD, both GRPO and SarcasmReasoner yield significantly higher paired accuracy than SFT after Holm correction ($p_{\mathrm{Holm}}=.037$ and $.040$); no pairwise classification difference is significant on MUStARD++, and SarcasmReasoner does not significantly outperform GRPO on either dataset. Importantly, GenRM is designed to improve evidence grounding apart from classification performance alone. We therefore examine its effects on reasoning quality through GenRM-based assessment and human evaluation in the following sections.

% These results suggest that GenRM-guided optimization provides an additional improvement in classification performance. Given the limited test-set sizes, these improvements should be interpreted cautiously (corresponding to a net increase of four correctly classified instances on MUStARD++ and four on MCSD, out of 181 and 406 test instances, respectively). More importantly, GenRM is designed to improve evidence grounding apart from classification performance alone. We therefore examine its effects on reasoning quality through GenRM-based assessment and independent human evaluation in the following sections.

Overall, these results demonstrate the complementary contributions of teacher reasoning, diverse SFT initialization, and reward-based alignment across both datasets. Although larger improvements are observed on MCSD, this difference may reflect dataset-specific characteristics.
% where zero-shot models struggle more with capturing dataset-specific sarcasm cues.

\begin{table}[t]
\centering
\caption{
GenRM-based assessment of generated evidence grounding on MUStARD++ and MCSD. Scores are normalized to $\{0,0.5,1\}$, averaged across test samples, and reported as percentages (\%). Avg. denotes the mean across the three modalities.
}
\label{tab:genrm_test_reasoning}

\setlength{\tabcolsep}{4pt}
\renewcommand{\arraystretch}{1.05}

\begin{tabular}{llcccc}
\toprule
\textbf{Dataset} &
\textbf{Model} &
\textbf{Text} &
\textbf{Audio} &
\textbf{Visual} &
\textbf{Avg.} \\
\midrule

\multirow{3}{*}{MUStARD++}
& SFT     & 82.87 & 85.36 & 48.62 & 72.28 \\
& GRPO    & 80.94 & 87.85 & 42.54 & 70.44 \\
& Ours & \textbf{87.57} & \textbf{92.82}
          & \textbf{88.40} & \textbf{89.59} \\

\midrule

\multirow{3}{*}{MCSD}
& SFT     & 89.53 & 86.21 & 79.93 & 85.22 \\
& GRPO    & 89.78 & 87.56 & 77.34 & 84.89 \\
& Ours & \textbf{94.09} & \textbf{96.31}
          & \textbf{93.47} & \textbf{94.62} \\

\bottomrule
\end{tabular}
\end{table}

\begin{figure}[t]
    \centering
    \includegraphics[width=\columnwidth]{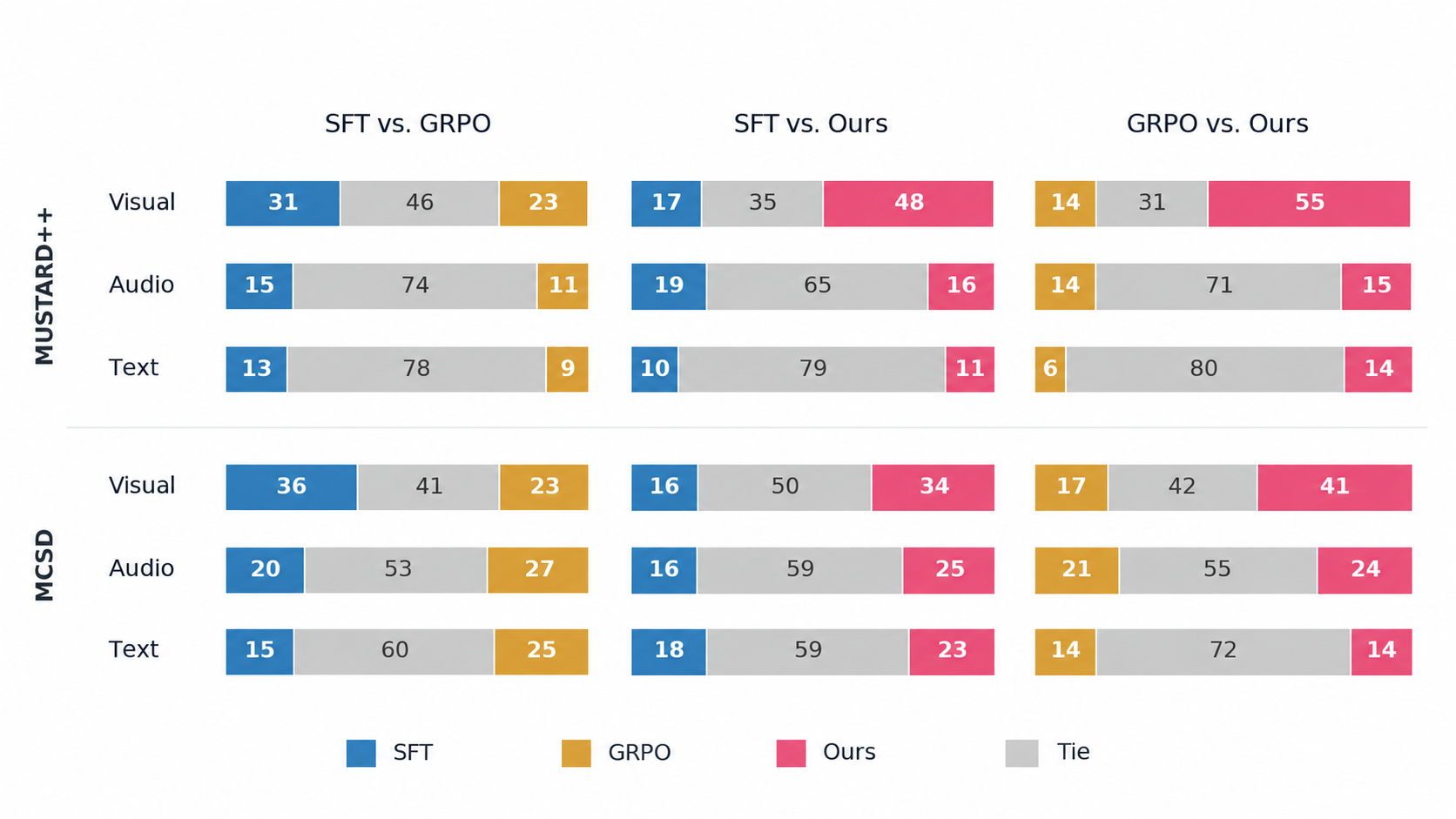}
    \caption{Pairwise preferences for modality-specific evidence grounding on MCSD and MUStARD++ (50 samples per dataset). Each stacked bar reports the percentages of preferences for the first system, ties, and preferences for the second system, averaged across the ratings of three annotators.}
    \label{fig:pairwise_preference}
\end{figure}
% Future: Interestingly, the approximately 14% textual grounding preference for our model is close to the 14.4% of sarcastic utterances categorized as embedded sarcasm in MUStARD++, where sarcasm can be identified from incongruity within the utterance itself \cite{ray-etal-2022-multimodal}. 

\noindent\textbf{GenRM-based Grounding Assessment.}
We further use GenRM to assess the grounding of generated evidence across the three post-training systems. As shown in Table~\ref{tab:genrm_test_reasoning}, standard GRPO slightly reduces the average GenRM-based grounding scores relative to SFT on both datasets, despite improving classification performance. In contrast, GenRM-guided GRPO increases the average grounding score from 70.44 to 89.59 on MUStARD++ and from 84.89 to 94.62 on MCSD. The largest improvement is observed in visual grounding, particularly on MUStARD++ (42.54 to 88.40). These results indicate that GenRM-guided optimization produces evidence that receives higher grounding scores from the reward model. Since the same GenRM is used for training and evaluation, we further assess reasoning quality based on evidence grounding through independent human evaluation.

\subsection{Human Evaluation}

Following the main comparison, we conduct a pairwise human preference evaluation on three representative post-training systems: SFT, SFT+GRPO, and SarcasmReasoner. These systems represent three post-training configurations of our framework and allow us to examine how supervised reasoning distillation, standard GRPO, and GenRM-guided alignment affect whether the generated reasoning is grounded in observable multimodal evidence, providing a complementary perspective to standard classification metrics.

Figure~\ref{fig:pairwise_preference} summarizes the pairwise preferences for modality-specific evidence grounding, averaged across the ratings from three annotators. On MUStARD++, the GenRM-guided model shows the clearest preference advantage in visual grounding, receiving 55.3\% of preferences compared with 14.0\% for standard GRPO and 48.0\% compared with 17.3\% for SFT. Both differences are statistically significant according to utterance-level exact sign tests (Holm-adjusted $p=0.0021$ and $p=0.0336$, respectively). Textual and acoustic comparisons are more balanced, with ties accounting for a substantial proportion of judgments.

A similar pattern emerges on MCSD. The GenRM-guided model receives 41.3\% of preferences compared with 16.7\% for standard GRPO and 34.0\% compared with 16.0\% for SFT in visual grounding. Both differences reach statistical significance according to utterance-level exact sign tests (Holm-adjusted $p=0.0319$ and $p=0.0499$, respectively). The differences are relatively modest for textual and acoustic evidence, where ties remain frequent.
Together with the improvements in classification performance, these findings suggest that GenRM provides a complementary reward signal that encourages more evidence-grounded explanations, particularly for visual evidence.

\begin{figure}[t]
    \centering
    \includegraphics[width=\columnwidth]{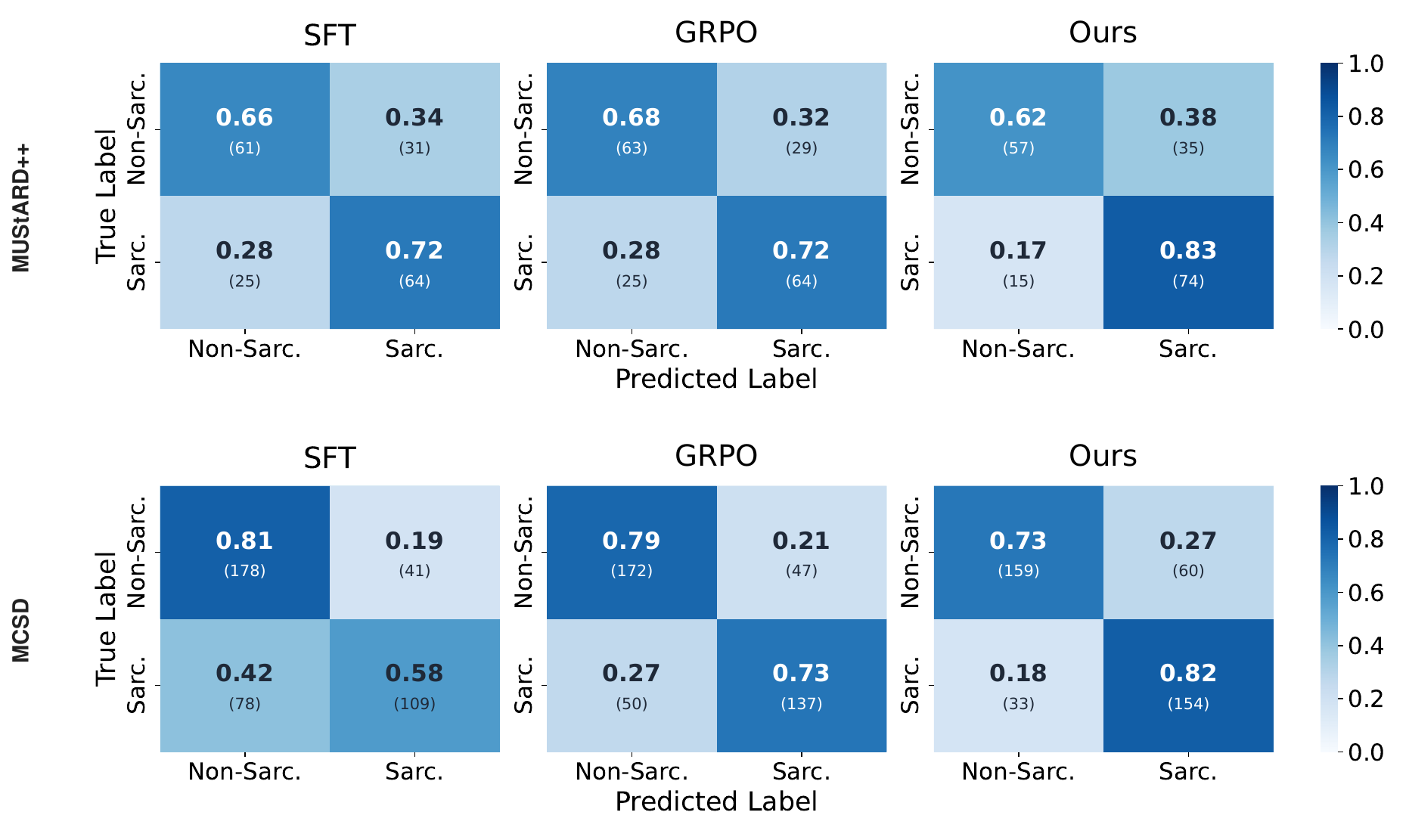}
    \caption{Row-normalized confusion matrices for different post-training methods on MUStARD++ and MCSD.}
    \label{fig:confusion_matrices}
\end{figure}

\begin{figure*}[t]
    \centering
    \includegraphics[width=\linewidth]
        {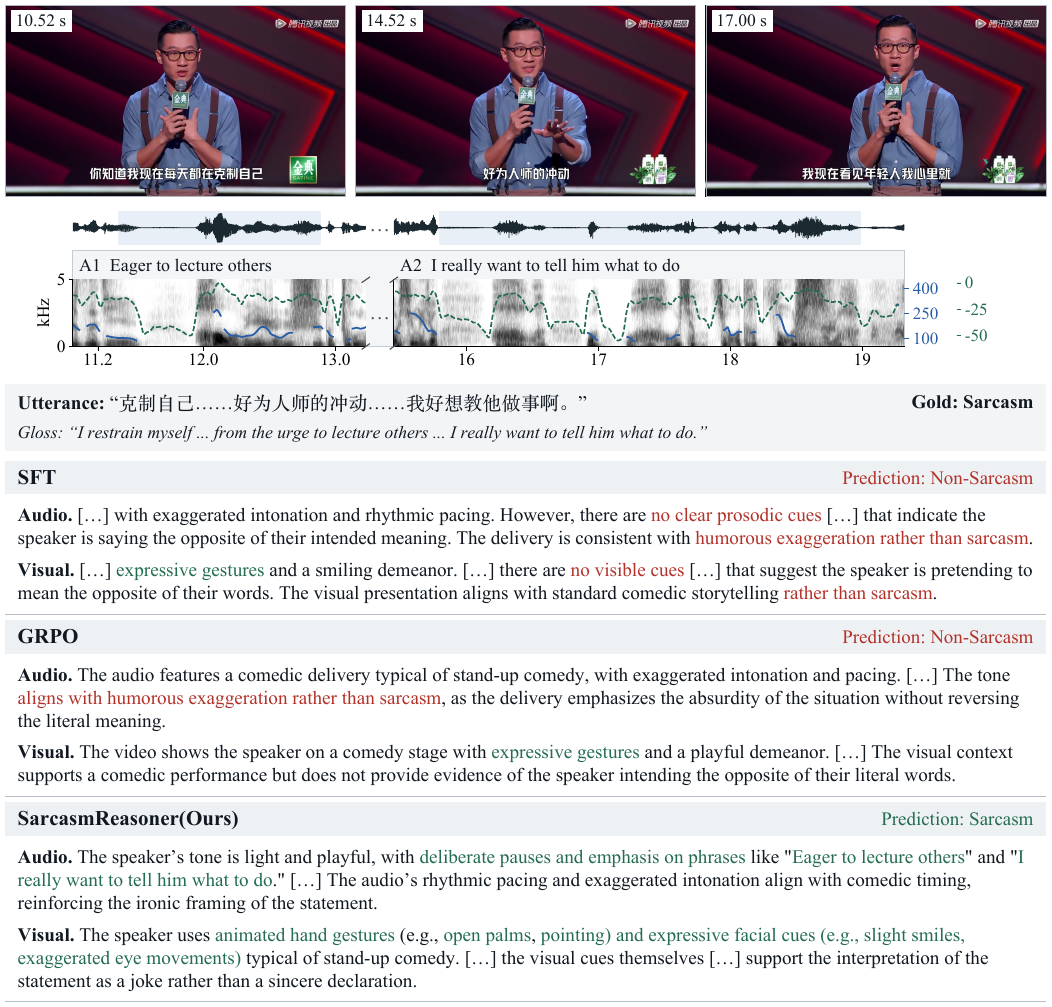}
    \caption{
        Qualitative comparison of generated evidence for a
        sarcastic MCSD utterance, highlighting differences
        in audio and visual reasoning.
    }
    \label{fig:case_study}
\end{figure*}
% This is a single illustrative case. The aggregate evidence for grounding is the human evaluation above.

\subsection{Error Analysis}
To complement the human evaluation of evidence grounding, we further analyze model errors at the label-decision level.
Using the same three representative post-training systems, we compare their row-normalized confusion matrices on MUStARD++ and MCSD in Figure~\ref{fig:confusion_matrices}.
This analysis allows us to examine how standard GRPO and GenRM-guided alignment change the model's tendency to confuse sarcastic and non-sarcastic samples.
The confusion matrices in Figure~\ref{fig:confusion_matrices} reveal different error patterns across the three post-training configurations. After SFT, the model exhibits distinct tendencies across datasets: on MUStARD++, it more frequently misclassifies non-sarcastic utterances as sarcastic, whereas on MCSD, it tends to miss genuinely sarcastic utterances. Adding standard GRPO produces relatively minor changes on MUStARD++, but substantially reduces false negatives on MCSD, improving the recognition of sarcastic utterances at the cost of a slight increase in false positives. Incorporating the GenRM reward further reduces false negatives on both datasets, indicating improved sensitivity to sarcastic cues. However, this improvement is accompanied by an increase in false positives, suggesting that the model becomes more likely to interpret non-sarcastic utterances as sarcastic. This trade-off highlights that improvements in sarcasm recognition do not necessarily translate into uniformly better classification across both classes. Together with the human evaluation results, these findings suggest that GenRM-guided alignment can encourage more evidence-grounded explanations while improving sarcasm recall, although it does not eliminate the ambiguity between sarcastic and non-sarcastic interpretations.

\noindent\textbf{Case Study.}
To illustrate how GenRM-guided alignment affects multimodal sarcasm reasoning, we examine a sarcastic utterance from MCSD (Figure~\ref{fig:case_study}). The speaker humorously describes trying to restrain his urge to lecture others while simultaneously admitting, ``I really want to tell him what to do.'' The contrast between his stated restraint and underlying desire creates a potential source of irony, further conveyed through his delivery and gestures.

Both SFT and SFT+GRPO incorrectly predict non-sarcasm. Although they recognize exaggerated intonation, rhythmic pacing, and expressive gestures, they interpret these cues as ordinary comedic exaggeration rather than sarcasm. Their explanations emphasize the absence of an explicit reversal of literal meaning, overlooking the ironic tension within the utterance. In contrast, SarcasmReasoner correctly identifies the utterance as sarcastic. Its explanation highlights deliberate pauses and emphasis on key phrases, together with animated gestures and facial expressions, interpreting these cues as supporting the speaker's playful and ironic framing rather than a sincere declaration.

This example illustrates that multimodal sarcasm reasoning depends not only on identifying relevant acoustic and visual cues, but also on interpreting how these cues contribute to the speaker's implied intent. GenRM-guided alignment can encourage more context-sensitive interpretations of multimodal evidence.

\begin{table}[t]
\centering
\small
\caption{
Cross-lingual performance under monolingual and bilingual training settings.
Scores are Macro-F1 (\%). Avg. denotes the unweighted average over
MUStARD++ and MCSD.
}
\label{tab:cross_lingual}
\resizebox{\columnwidth}{!}{
\begin{tabular}{llccc}
\toprule
\textbf{Training}
& \textbf{Data}
& \textbf{MUStARD++}
& \textbf{MCSD}
& \textbf{Avg.} \\
\midrule

\multicolumn{5}{l}{\textit{Monolingual SFT}} \\

SFT
& EN only
& \textbf{69.05}
& 59.68
& 64.37 \\

SFT
& ZH only
& 63.71
& \textbf{69.82}
& \textbf{66.77} \\

\midrule

\multicolumn{5}{l}{\textit{Bilingual SFT}} \\

SFT
& Downsample
& 62.40
& \textbf{72.20}
& \textbf{67.30} \\

SFT
& Oversample
& \textbf{65.19}
& 68.95
& 67.07 \\

SFT
& Pooled
& 64.61
& 69.56
& 67.09 \\

\midrule

\multicolumn{5}{l}{\textit{Bilingual post-training (pooled)}} \\

GRPO
& Pooled
& 64.46
& 74.28
& 69.37 \\

Ours
& Pooled
& \textbf{65.18}
& \textbf{76.84}
& \textbf{71.01} \\

\bottomrule
\end{tabular}
}
\end{table}

\subsection{Cross-lingual Generalization} 

We further evaluate whether sarcasm reasoning learned from one language transfers to the other and whether bilingual training improves performance across English and Chinese.
Table~\ref{tab:cross_lingual} summarizes the results under monolingual SFT, bilingual SFT, and bilingual post-training settings.

Among the monolingual SFT models, in-language training achieves stronger performance on the corresponding test sets: EN-only training achieves 69.05\% Macro-F1 on MUStARD++, while ZH-only training achieves 69.82\% on MCSD.
Cross-lingual transfer is possible but asymmetric: the ZH-only model achieves 63.71\% on MUStARD++, while the EN-only model reaches 59.68\% on MCSD, corresponding to performance drops of 5.34 and 10.14 points relative to in-language training.
This suggests that reasoning patterns learned from Chinese data may generalize more effectively to English than vice versa, although in-language supervision remains important for capturing dataset-specific pragmatic cues.
% This suggests that the reasoning patterns learned from MCSD may provide broader pragmatic supervision, possibly due to the larger scale or greater diversity of the Chinese training set. 

Bilingual SFT provides a slightly better average trade-off, although the effects of data balancing vary across datasets.
Downsampling achieves the highest average Macro-F1 (67.30\%) and the best performance on MCSD (72.20\%), while oversampling performs best on MUStARD++ (65.19\%) among bilingual SFT settings.
However, the average differences among the three bilingual strategies are small (67.07--67.30\%), suggesting that data balancing does not consistently improve performance across both datasets.

Bilingual post-training further improves the average Macro-F1 from 67.09\% with pooled SFT to 69.37\% with GRPO and 71.01\% with SarcasmReasoner.
The improvement is particularly substantial on MCSD, where Macro-F1 increases from 69.56\% to 76.84\%. Overall, these results suggest that bilingual optimization does not necessarily yield balanced improvements across languages. In particular, bilingual SarcasmReasoner performs substantially below its English-only counterpart on MUStARD++ (65.18\% vs. 72.13\%).

\section{Discussion}

\subsection{From Multimodal Prediction to Evidence-Grounded Reasoning}
The results highlight the benefits of task-specific post-training for multimodal sarcasm detection compared with zero-shot inference.
% Model size: 7B vs. 30B
% Training strategy: post-trained vs. zero-shot
While larger omni-modal models demonstrate some zero-shot capability, their performance remains inconsistent, particularly on the more challenging MCSD dataset. In contrast, the post-trained 7B model consistently outperforms its larger zero-shot teacher. This finding indicates that targeted supervision and alignment are important for learning the subtle interactions among lexical meaning, prosody, and facial behavior that characterize sarcastic communication. 

Classification performance and evidence grounding do not necessarily improve in parallel. While standard GRPO improves Macro-F1 over SFT on both datasets, the GenRM-based assessment shows no corresponding improvement in evidence grounding. In contrast, GenRM-guided optimization provides additional gains in Macro-F1 over standard SFT+GRPO, while human evaluation also shows a preference for GenRM-guided explanations on both datasets, particularly in visual grounding. Taken together, these findings highlight the importance of evaluating evidence grounding alongside classification performance, as improvements in predictive accuracy alone do not guarantee more reliable explanations.

The human evaluation provides further evidence that GenRM-guided optimization produces better-grounded explanations, particularly for visual evidence on MUStARD++. However, the error analysis also reveals an increased false-positive rate for SarcasmReasoner, indicating a greater tendency to interpret non-sarcastic utterances as sarcastic. This highlights a trade-off between evidence grounding and classification behavior, suggesting that improvements in explanation quality do not necessarily translate into more reliable sarcasm predictions.
% The error analysis further suggests that optimizing answer correctness alone may change the model's classification behavior without ensuring that its explanations are supported by multimodal evidence. In contrast, GenRM-guided optimization encourages explanations that are more consistent with observable evidence, rather than relying on plausible but unsupported interpretations.
% % This indicates that optimizing for answer correctness alone may refine the model’s decision boundary while leaving unsupported multimodal interpretations largely unaddressed. 

\subsection{Complementary Roles of Distillation and Reward Alignment}

The ablation results suggest that distillation and reward-based optimization play complementary roles in SarcasmReasoner. 
Diverse-8 achieves the strongest SFT performance among the evaluated strategies, suggesting that learning from multiple reasoning trajectories is a promising direction for multimodal sarcasm detection. %  likely because sarcasm can be supported by different combinations of lexical, acoustic, visual, and contextual evidence. 
As the strategies also differ in data size and training duration, the specific contribution of trajectory diversity remains to be established.

Policy optimization is most effective once such a task-relevant reasoning prior has been established. GRPO applied directly to the zero-shot model performs poorly, whereas initializing from the SFT model leads to substantial improvements. However, standard GRPO can improve classification performance without consistently improving explanation quality, showing that answer-level rewards alone provide insufficient constraints on multimodal reasoning. GenRM complements the accuracy reward by encouraging more reliable and better-grounded justifications.

% The reward-model ablation further shows that more complex reasoning rewards do not necessarily yield better policies, with performance varying across datasets. Future work could therefore explore finer-grained reward signals that evaluate individual textual, acoustic, and visual claims rather than assigning a single score to the entire reasoning trajectory.

\subsection{Language and Dataset Effects}

The larger gains on MCSD should not be attributed to language alone. MUStARD++ and MCSD also differ in genre, dataset size, utterance duration, speaker behavior, and pragmatic cues. In particular, MCSD contains longer utterances, which may make it harder for the model to identify which textual, acoustic, and visual cues are relevant to sarcasm. The lower zero-shot performance and larger post-training gains on MCSD may therefore reflect both language and dataset differences. 

Transfer between the two datasets is partial and asymmetric, with monolingual models performing better on their respective datasets. As the datasets differ in language, genre (sitcom vs. stand-up comedy), and training-set size, this asymmetry cannot be attributed to language alone. Bilingual training does not fully close the performance gap, with the three sampling strategies yielding similar average performance. Although bilingual post-training improves performance on MCSD, it remains less effective than English-only training on MUStARD++. Separating transferable pragmatic cues from language- and genre-specific ones would require datasets that vary language while controlling for genre and other dataset characteristics.

% "Transfer between the two datasets is partial and asymmetric: SFT on MCSD loses 5.34 points on MUStARD++ relative to in-language training, while SFT on MUStARD++ loses 10.14 points on MCSD. As the datasets differ in language, genre (sitcom vs. stand-up), and training-set size, this asymmetry cannot be attributed only to language. Bilingual training does not close the gap. The three bilingual sampling strategies yield similar average performance (67.07–67.30%), and pooled post-training improves MCSD while leaving MUStARD++ (65.18%)  below the English-only SarcasmReasoner (72.13%). Separating transferable pragmatic cues from language- and genre-specific ones would require data that varies language while keeping genre constant."

\subsection{Limitations and Future Directions}

This study has several limitations. First, evaluation is limited to MUStARD++ and MCSD. This is partly due to the scarcity of spoken multimodal sarcasm datasets with aligned text, audio, and video. Sarcasm annotation is also costly and subjective, making large-scale dataset construction difficult. Future work should therefore focus on building larger and more diverse datasets, potentially using scalable annotation methods with human verification.
Second, the pairwise human evaluation is limited to 50 samples per dataset and three annotators. 
Moreover, non-native English annotators may be less sensitive to subtle prosodic cues in MUStARD++.
A larger study with more annotators would provide more reliable estimates of modality-specific grounding preferences and help determine whether the observed differences generalize beyond the evaluated samples. 
Third, a plausible explanation is not necessarily faithful to the model's actual reasoning. Counterfactual tests, such as removing or replacing individual modalities, could provide stronger evidence that the model relies on the cues described in its explanation.
Moreover, GenRM relies on automatically derived grounding scores, and the relatively low agreement between visual judges may introduce noise or selection bias into the supervision. Since the same GenRM is used for reward optimization and automatic grounding assessment, the latter should be interpreted as a model-based diagnostic rather than independent validation. Correctness-gated GenRM rewards introduce an additional accuracy signal, potentially confounding the attribution of classification gains over standard GRPO.
Furthermore, the SFT strategies differ in training data size and duration, making it difficult to isolate the contribution of trajectory diversity to performance improvements. The experiments use a single student backbone, while both the training trajectories and reasoning reward are derived from the teacher model. This may introduce teacher-specific biases. 
Future work should evaluate different model families and develop reward models that can directly verify textual, acoustic, and visual evidence. Finally, the training experiments are conducted with a single run per configuration due to limited computational resources.

% x

\section{Conclusion}

This work introduced SarcasmReasoner, a three-stage post-training framework for evidence-grounded text-audio-visual sarcasm reasoning. The framework combines diverse teacher trajectory sampling, dual-track distillation of correct and incorrect trajectories, and GRPO with accuracy, format, and GenRM-based grounding rewards. On MUStARD++ and MCSD, the post-trained 7B model substantially outperforms its zero-shot backbone and the larger zero-shot teacher, reaching Macro-F1 scores of 72.13\% and 77.09\%, respectively. 
Stage-wise ablations highlight the benefits of Diverse-8 SFT initialization and subsequent reward-based optimization.
% Stage-wise ablations show that teacher reasoning, trajectory diversity, SFT initialization, and reward alignment make complementary contributions.
Pairwise human evaluation further suggests improvements in modality-specific evidence grounding on both datasets, particularly for visual evidence. Cross-lingual results show partial but asymmetric transfer and suggest that language balancing does not consistently improve performance. Taken together, these findings support treating multimodal sarcasm detection as both a prediction problem and an evidence-grounding problem, while motivating further work on faithful and reliable multimodal reasoning.

% \section*{Acknowledgments}

\bibliographystyle{IEEEtran}
\bibliography{references}

\vfill
\clearpage
\appendix

% \section{Additional Experimental Details}
% \label{app:additional_details}

% \noindent\textbf{Appendix contents.}\\[-1pt]
% {\small
% \noindent\ref*{app:datasets}~Dataset Statistics\\
% \ref*{app:sft_data}~Teacher Trajectory Filtering and SFT Data\\
% \ref*{app:judge_data}~Evidence-Quality Supervision\\
% \ref*{app:training_setup}~Model Selection and Training Configuration\\
% \ref*{app:prompts}~Prompt Templates
% }
% \medskip

\subsection{Ethical Considerations}
\label{app:ethical_considerations}

\noindent\textbf{Data Use and Privacy.}
We use two publicly released research datasets, MUStARD++ \cite{ray-etal-2022-multimodal} and MCSD \cite{gao2025multimodal}, containing audiovisual excerpts from television sitcoms and stand-up comedy. Although the recordings originate from publicly broadcast material, they feature identifiable performers and may be subject to copyright and privacy considerations. We use these datasets solely for academic research, acknowledge their original sources, and do not redistribute the original audiovisual recordings. Our study does not attempt to identify, track, or profile individuals appearing in the videos.

\noindent\textbf{Human Evaluation.}
The human preference evaluation was conducted by three annotators per dataset. For each dataset, the annotators independently assessed 50 samples through blinded pairwise comparisons of model-generated explanations. For each sample, annotators were presented with two anonymized modality-specific evidence sections. They saw only the evidence sections for the modality being assessed; Gold sarcasm labels, and the `<integration>` and `<answer>` sections were withheld. 
Textual, audio, and visual evidence were evaluated separately, with ties permitted when neither explanation was preferred. The evaluation focused on the quality and grounding of generated evidence rather than the personal characteristics of the performers. We report inter-annotator agreement to account for the subjectivity of these judgments.

\noindent\textbf{Limitations and Responsible Use.}
Sarcasm recognition involves inferring a speaker's communicative intent from linguistic, acoustic, and visual cues. Such inferences are inherently uncertain and may vary across languages, cultures, and individuals. Models trained on scripted or performed speech may not generalize reliably to spontaneous interactions or speakers with different expressive styles. Moreover, generated explanations may contain unsupported claims and should not be treated as definitive evidence of a speaker's intentions or emotional state.

SarcasmReasoner is intended as a research tool for multimodal language understanding and has not been validated for making decisions about individuals. Related applications may also be subject to legal restrictions. For example, Article~5(1)(f) of the EU AI Act prohibits certain uses of AI systems to infer emotions in workplaces and educational institutions, subject to exceptions for medical or safety purposes.\footnote{\url{https://eur-lex.europa.eu/eli/reg/2024/1689/oj}} Whether this provision applies to a particular sarcasm-detection application depends on its intended purpose and use.

% \subsection{Statistical Significance and Experimental Variability}
% Due to the substantial computational cost of multimodal post-training, conducting multiple independent runs for every experimental configuration is computationally expensive. 
% To assess the stability of our results, we repeated a representative GRPO post-training pipeline using different random seeds and the observed performance variation is very small across runs.
% Although we do not report error bars for all experimental configurations, this result provides preliminary evidence of the stability of GRPO-based post-training.

\subsection{Statistical Significance}

Due to computational cost, each reported configuration is trained once, and we therefore do not estimate variability across training seeds. The analyses below instead assess paired differences between systems on the same test instances.

\noindent\textbf{Paired classification comparisons.} Table~\ref{tab:paired_classification_tests} reports all pairwise comparisons among Diverse-8 SFT, standard GRPO, and SarcasmReasoner. McNemar's test evaluates paired correctness, whereas the bootstrap intervals describe Macro-F1 differences. After Holm correction across all six tests, GRPO and SarcasmReasoner each yield significantly higher paired correctness than SFT on MCSD; the remaining classification differences are not statistically significant.

\begin{table*}[t]
\centering
\caption{Paired classification comparisons. $\Delta$ Macro-F1 is the Macro-F1 difference in percentage points from model A to model B. Here, $b$ counts instances correct only under A and $c$ those correct only under B. Raw two-sided exact McNemar $p$ values are corrected jointly using Holm's procedure. CIs are pointwise paired-percentile-bootstrap 95\% intervals for $\Delta$ Macro-F1 based on 100,000 resamples.}
\label{tab:paired_classification_tests}
\small
\setlength{\tabcolsep}{5pt}
\begin{tabular*}{\textwidth}{@{\extracolsep{\fill}}llrrrrrl@{}}
\toprule
\textbf{Dataset} & \textbf{A $\rightarrow$ B} & \textbf{$\Delta$ Macro-F1} & \textbf{$b$} & \textbf{$c$} & \textbf{Raw $p$} & \textbf{Holm $p$} & \textbf{95\% CI} \\
\midrule
\multirow{3}{*}{MUStARD++}
& SFT $\rightarrow$ GRPO             & +1.11 & 10 & 12 & .832 & 1.000 & $[-3.88, 6.13]$ \\
& SFT $\rightarrow$ SarcasmReasoner  & +3.08 & 10 & 16 & .327 & 1.000 & $[-2.43, 8.68]$ \\
& GRPO $\rightarrow$ SarcasmReasoner & +1.97 & 9  & 13 & .523 & 1.000 & $[-3.04, 7.03]$ \\
\midrule
\multirow{3}{*}{MCSD}
& SFT $\rightarrow$ GRPO             & +6.11 & 19 & 41 & .006 & \textbf{.037} & $[2.35, 9.94]$ \\
& SFT $\rightarrow$ SarcasmReasoner  & +7.27 & 32 & 58 & .008 & \textbf{.040} & $[2.76, 11.84]$ \\
& GRPO $\rightarrow$ SarcasmReasoner & +1.16 & 24 & 28 & .678 & 1.000 & $[-2.31, 4.65]$ \\
\bottomrule
\end{tabular*}
\end{table*}

\subsection{Dataset Statistics}
\label{app:datasets}

Table~\ref{tab:dataset_statistics} summarizes the training, validation, and test splits used in our experiments on MUStARD++ and MCSD. The datasets are approximately balanced overall, while MCSD has longer utterances and substantially more video material. Text length is measured in words for MUStARD++ and in Chinese characters for MCSD.

\begin{table*}[h]
\centering
\caption{Dataset statistics for MUStARD++ and MCSD. Text length is measured in
words for MUStARD++ and Chinese characters for MCSD.}
\label{tab:dataset_statistics}
\small
\setlength{\tabcolsep}{4pt}
\renewcommand{\arraystretch}{1.07}
\begin{tabular*}{\textwidth}{@{\extracolsep{\fill}}llrrrrrr@{}}
\toprule
\textbf{Dataset} & \textbf{Split} & \textbf{\# Samples} &
\textbf{Sarcasm} & \textbf{Non-sarcasm} &
\shortstack{\textbf{Mean text}\\\textbf{length}} &
\shortstack{\textbf{Mean video}\\\textbf{duration}} &
\shortstack{\textbf{Total}\\\textbf{duration}} \\
\midrule
\multirow{4}{*}{MUStARD++}
& Train & 841 & 415 & 426 & 11.49 words & 4.60 s & 1.07 h \\
& Valid & 180 & 97 & 83 & 11.08 words & 4.73 s & 0.24 h \\
& Test & 181 & 89 & 92 & 11.10 words & 4.51 s & 0.23 h \\
& \textbf{Total} & \textbf{1,202} & \textbf{601} & \textbf{601} &
\textbf{11.37 words} & \textbf{4.60 s} & \textbf{1.54 h} \\
\midrule
\multirow{4}{*}{MCSD}
& Train & 1,893 & 936 & 957 & 64.09 chars & 14.22 s & 7.48 h \\
& Valid & 406 & 221 & 185 & 62.04 chars & 13.77 s & 1.55 h \\
& Test & 406 & 187 & 219 & 62.76 chars & 13.93 s & 1.57 h \\
& \textbf{Total} & \textbf{2,705} & \textbf{1,344} & \textbf{1,361} &
\textbf{63.58 chars} & \textbf{14.11 s} & \textbf{10.60 h} \\
\bottomrule
\end{tabular*}
\end{table*}

\subsection{Teacher Trajectory Filtering and SFT Data}
\label{app:sft_data}

\noindent\textbf{Teacher sampling and output structure.} For each multimodal input, the Qwen3-Omni-30B-A3B-Thinking teacher generates eight independently sampled trajectories using temperature $T=0.6$ and $\mathrm{top}\text{-}p=0.95$. Each output is required to contain exactly one nonempty instance of the five tagged elements \texttt{<text\_evidence>}, \texttt{<audio\_evidence>}, \texttt{<visual\_evidence>}, \texttt{<integration>}, and \texttt{<answer>}, with corresponding closing tags. The answer must be either \texttt{Sarcasm} or \texttt{Non-Sarcasm}. Additional free-form teacher thinking is not included in the student training targets. The teacher prompt is provided in Table~\ref{tab:sarcasm_reasoning_prompt}.

\noindent\textbf{SFT trajectory filtering.} Track~A retains a trajectory only when (i) its final answer matches the gold label, (ii) all five elements are structurally complete and nonempty, and (iii) generation terminates normally. Table~\ref{tab:sft_rejections} reports the mutually exclusive rejection categories for greedy decoding and eight-sample generation. These filtering counts refer to SFT construction; Track~B also retains structurally complete trajectories with incorrect predictions for grounding supervision.

\begin{table}[h]
\centering
\caption{Teacher-generation filtering on the training splits.
G denotes greedy decoding and 8 denotes eight-sample generation.
All entries are trajectory counts.}
\label{tab:sft_rejections}
\small
\setlength{\tabcolsep}{3pt}
\begin{tabular*}{\columnwidth}{@{\extracolsep{\fill}}lrrrr@{}}
\toprule
& \multicolumn{2}{c}{\textbf{MUStARD++}} &
  \multicolumn{2}{c}{\textbf{MCSD}} \\
\cmidrule(lr){2-3}\cmidrule(l){4-5}
\textbf{Count} & \textbf{G} & \textbf{8} & \textbf{G} & \textbf{8} \\
\midrule
Generated & 841 & 6,728 & 1,893 & 15,144 \\
Malformed & 48 & 253 & 38 & 147 \\
Wrong prediction & 243 & 2,001 & 643 & 5,300 \\
Non-stop finish & 0 & 27 & 0 & 81 \\
\midrule
\textbf{Rejected} & \textbf{291} & \textbf{2,281} &
\textbf{681} & \textbf{5,528} \\
\bottomrule
\end{tabular*}
\end{table}

\noindent\textbf{Trajectory selection strategies.} Greedy decoding retains the single deterministic trajectory when it passes the SFT filter. Best-of-8 samples eight candidates and retains the eligible trajectory with the highest teacher-reported mean token log-probability. Diverse-8 retains all eligible sampled trajectories. Appendix~\ref{app:sft_data} reports the number of training trajectories, the fraction of source utterances covered, and the mean number of retained trajectories per covered source. Best-of-8 increases source coverage relative to greedy decoding, while Diverse-8 provides multiple trajectories for the same covered utterances.

\begin{table}[!htbp]
\centering
\caption{Usable SFT training data. Coverage is the fraction of source
utterances represented by at least one retained trajectory; Traj./src.
is the mean number of retained trajectories per covered source.}
\label{tab:sft_train_stats}
\small
\setlength{\tabcolsep}{4pt}
\begin{tabular*}{\columnwidth}{@{\extracolsep{\fill}}lrrr@{}}
\toprule
\textbf{Strategy} & \textbf{SFT ex.} & \textbf{Coverage} &
\textbf{Traj./src.} \\
\midrule
\multicolumn{4}{@{}l}{\emph{MUStARD++}} \\
Greedy & 550 & 65.4\% & 1.00 \\
Best-of-8 & 757 & 90.0\% & 1.00 \\
Diverse-8 & \textbf{4,447} & 90.0\% & \textbf{5.87} \\
\addlinespace[3pt]
\multicolumn{4}{@{}l}{\emph{MCSD}} \\
Greedy & 1,212 & 64.0\% & 1.00 \\
Best-of-8 & 1,669 & 88.2\% & 1.00 \\
Diverse-8 & \textbf{9,616} & 88.2\% & \textbf{5.76} \\
\bottomrule
\end{tabular*}
\end{table}

\subsection{GenRM Supervision and Judge Agreement}
\label{app:judge_data}

\noindent\textbf{Modality-specific assessment.} Each teacher-generated trajectory eligible for Track~B is assessed independently by two judge models per modality, as specified in the experimental setup. Text judges inspect the transcript and generated text evidence. Audio judges inspect the audio and generated audio evidence, with the transcript available as linguistic context. Visual judges inspect ordered video frames and generated visual evidence, also using the transcript as linguistic context. The gold label, final prediction, and integration evidence are not supplied as separate judging inputs. The three-level scores reflect the extent to which observations and interpretations are supported by the designated modality. Table~\ref{tab:individual_judges} reports the individual judge score distributions.

\begin{table}[!htbp]
\centering
\caption{Individual modality-specific judge scores on training
trajectories. S1--S3 are the percentage distributions for unreliable,
partially reliable, and reliable evidence, respectively; Mean is the
mean three-level score. All outputs are parsed after retry recovery.}
\label{tab:individual_judges}
\footnotesize
\setlength{\tabcolsep}{2pt}
\begin{tabular*}{\columnwidth}{@{\extracolsep{\fill}}lrrrr@{}}
\toprule
\textbf{Judge} & \textbf{S1} & \textbf{S2} & \textbf{S3} &
\textbf{Mean} \\
\midrule
\multicolumn{5}{@{}l}{\emph{MUStARD++ -- Text}} \\
Gemma-3-12B & 15.4 & 14.1 & 70.5 & 2.551 \\
DeepSeek-V2-Lite & 16.0 & 3.1 & 80.8 & 2.648 \\
\multicolumn{5}{@{}l}{\emph{MUStARD++ -- Audio}} \\
Step-Audio-2-mini & 2.1 & 31.8 & 66.1 & 2.640 \\
Phi-4-Multimodal & 4.6 & 37.4 & 57.9 & 2.533 \\
\multicolumn{5}{@{}l}{\emph{MUStARD++ -- Visual}} \\
InternVL3.5-8B & 19.3 & 22.0 & 58.7 & 2.395 \\
Molmo2-8B & 38.8 & 41.9 & 19.3 & 1.805 \\
\addlinespace[3pt]
\multicolumn{5}{@{}l}{\emph{MCSD -- Text}} \\
Gemma-3-12B & 9.2 & 21.2 & 69.6 & 2.604 \\
DeepSeek-V2-Lite & 12.8 & 1.5 & 85.7 & 2.729 \\
\multicolumn{5}{@{}l}{\emph{MCSD -- Audio}} \\
Step-Audio-2-mini & 0.4 & 23.3 & 76.3 & 2.759 \\
Phi-4-Multimodal & 1.9 & 30.0 & 68.2 & 2.663 \\
\multicolumn{5}{@{}l}{\emph{MCSD -- Visual}} \\
InternVL3.5-8B & 3.0 & 15.6 & 81.4 & 2.784 \\
Molmo2-8B & 24.5 & 49.3 & 26.2 & 2.017 \\
\bottomrule
\end{tabular*}
\end{table}

\noindent\textbf{Consensus and masking.} For modality $m\in\mathcal{M}=\{\mathrm{text},\mathrm{audio},\mathrm{visual}\}$, let $s_m^{(1)}$ and $s_m^{(2)}$ denote the independently assigned judge scores. 
Because the three-level rubric is ordinal, we report quadratically weighted Cohen's $\kappa$ ($\kappa_w$) to measure chance-corrected agreement.
We accept a target score only when both judges agree on a valid value:
\begin{equation}
\mu_m=\mathbb{I}\!\left[s_m^{(1)}=s_m^{(2)}\in\{1,2,3\}\right].
\end{equation}
If $\mu_m=1$, the agreed score is used as the target $s_m$; otherwise, the corresponding score target is masked during training. A trajectory is retained when at least one modality has an agreed score. Thus, missing consensus does not by itself imply that the generated evidence is unreliable. Table~\ref{tab:genrm_supervision} reports exact agreement rates and the distribution of accepted scores, conditional on agreement.

\begin{table}[!htbp]
\centering
\caption{Dual-judge agreement and GenRM supervision on training trajectories (\%). Agree. is the exact score agreement rate, and $\kappa_w$ is quadratically weighted Cohen's kappa computed over all valid paired scores before consensus masking. S1-S3 are computed only among trajectories with exact agreement. Avg. is an unweighted mean across modalities.}
\label{tab:genrm_supervision}
\small
\setlength{\tabcolsep}{3pt}
\begin{tabular*}{\columnwidth}{@{\extracolsep{\fill}}lrrrrr@{}}
\toprule
\textbf{Modality} & \textbf{Agree.} & \textbf{$\kappa_w$} & \textbf{S1} & \textbf{S2} & \textbf{S3} \\
\midrule
\multicolumn{6}{@{}l}{\emph{MUStARD++}} \\
Text   & 75.6 & 0.523 & 11.6 & 1.6  & 86.8 \\
Audio  & 76.9 & 0.614 & 2.4  & 30.2 & 67.4 \\
Visual & 43.2 & 0.489 & 40.7 & 16.2 & 43.1 \\
\textbf{Avg.} & \textbf{65.2} & -- & \textbf{18.2} & \textbf{16.0} & \textbf{65.8} \\
\addlinespace[3pt]
\multicolumn{6}{@{}l}{\emph{MCSD}} \\
Text   & 73.3 & 0.518 & 7.5 & 1.0  & 91.6 \\
Audio  & 81.7 & 0.598 & 0.5 & 21.4 & 78.0 \\
Visual & 32.0 & 0.264 & 8.0 & 10.5 & 81.6 \\
\textbf{Avg.} & \textbf{62.4} & -- & \textbf{5.3} & \textbf{11.0} & \textbf{83.7} \\
\bottomrule
\end{tabular*}
\end{table}

\noindent\textbf{GenRM score prediction.} Qwen2.5-Omni-3B is fine-tuned to predict grounding scores from the original transcript, audio, ordered video frames, and the generated text, audio, and visual evidence. The gold sarcasm label, final prediction, and integration section are excluded. The target output is the compact JSON object \texttt{\{"text":X,"audio":Y,"visual":Z\}}, with each value in $\{1,2,3\}$. Only agreed modality-score targets contribute to the training loss; disagreeing targets are masked. At inference time, GenRM predicts all three modality scores. The shared judging rubric is reproduced in Table~\ref{tab:judge_prompt}.

\subsection{Human Evaluation Protocol}
\label{app:human_protocol}

\begin{figure*}[p]
\centering
\includegraphics[width=0.8\textwidth]{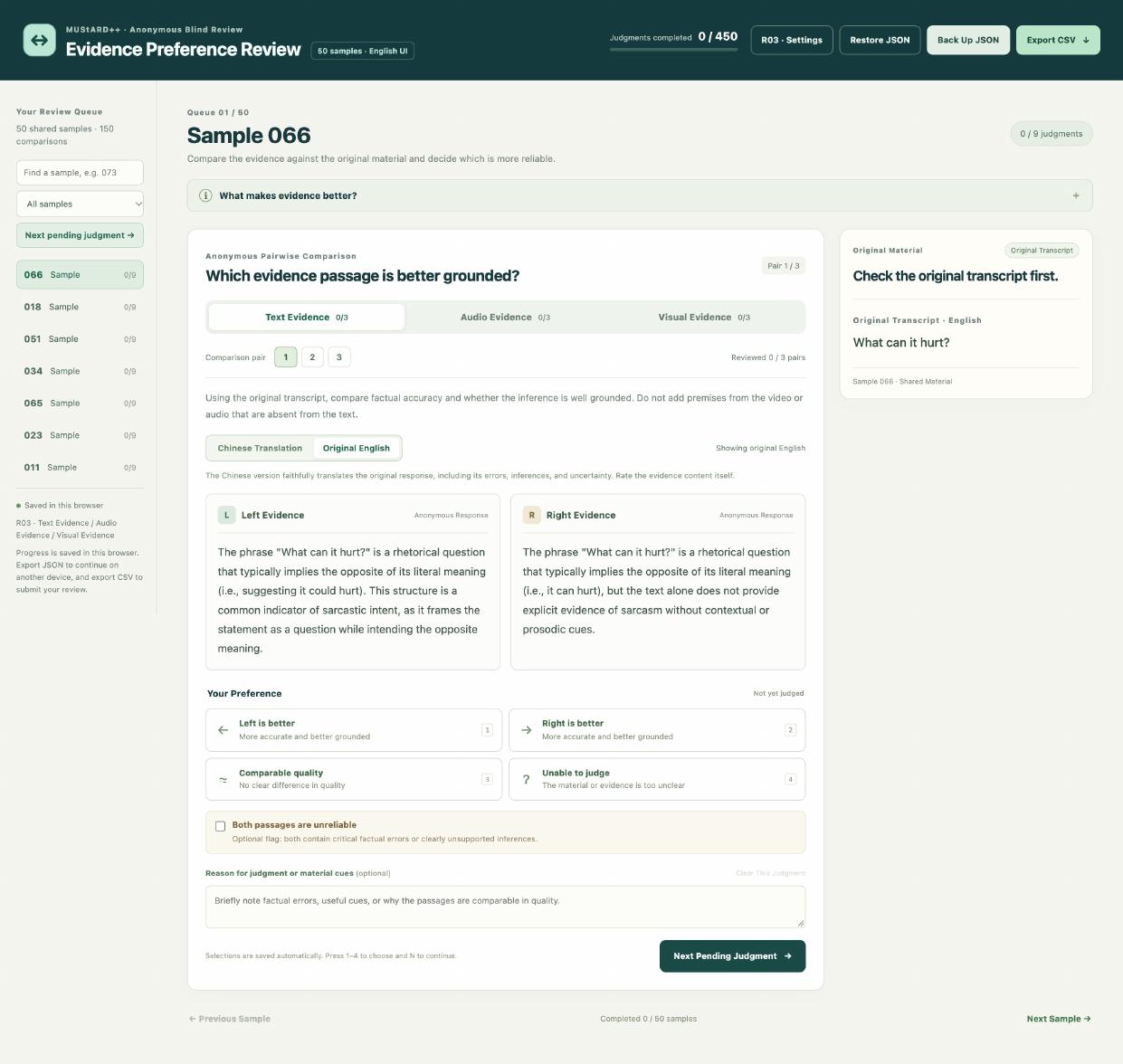}
\caption{Human evaluation interface for textual evidence grounding.}
\label{fig:human_eval_text}
\end{figure*}

\begin{figure*}[p]
\centering
\includegraphics[width=0.8\textwidth]{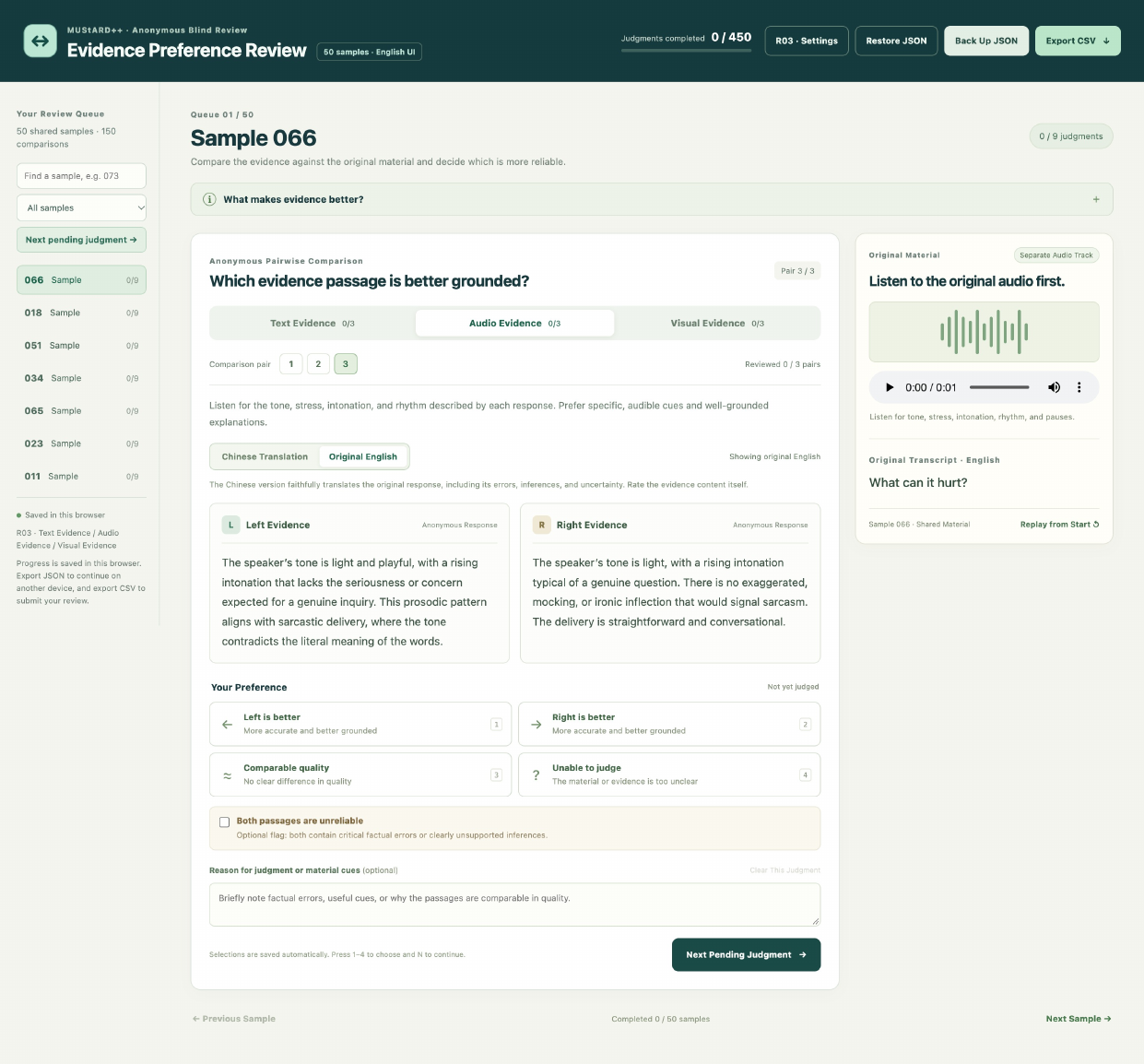}
\caption{Human evaluation interface for acoustic evidence grounding.}
\label{fig:human_eval_audio}
\end{figure*}

\begin{figure*}[p]
\centering
\includegraphics[width=0.8\textwidth]{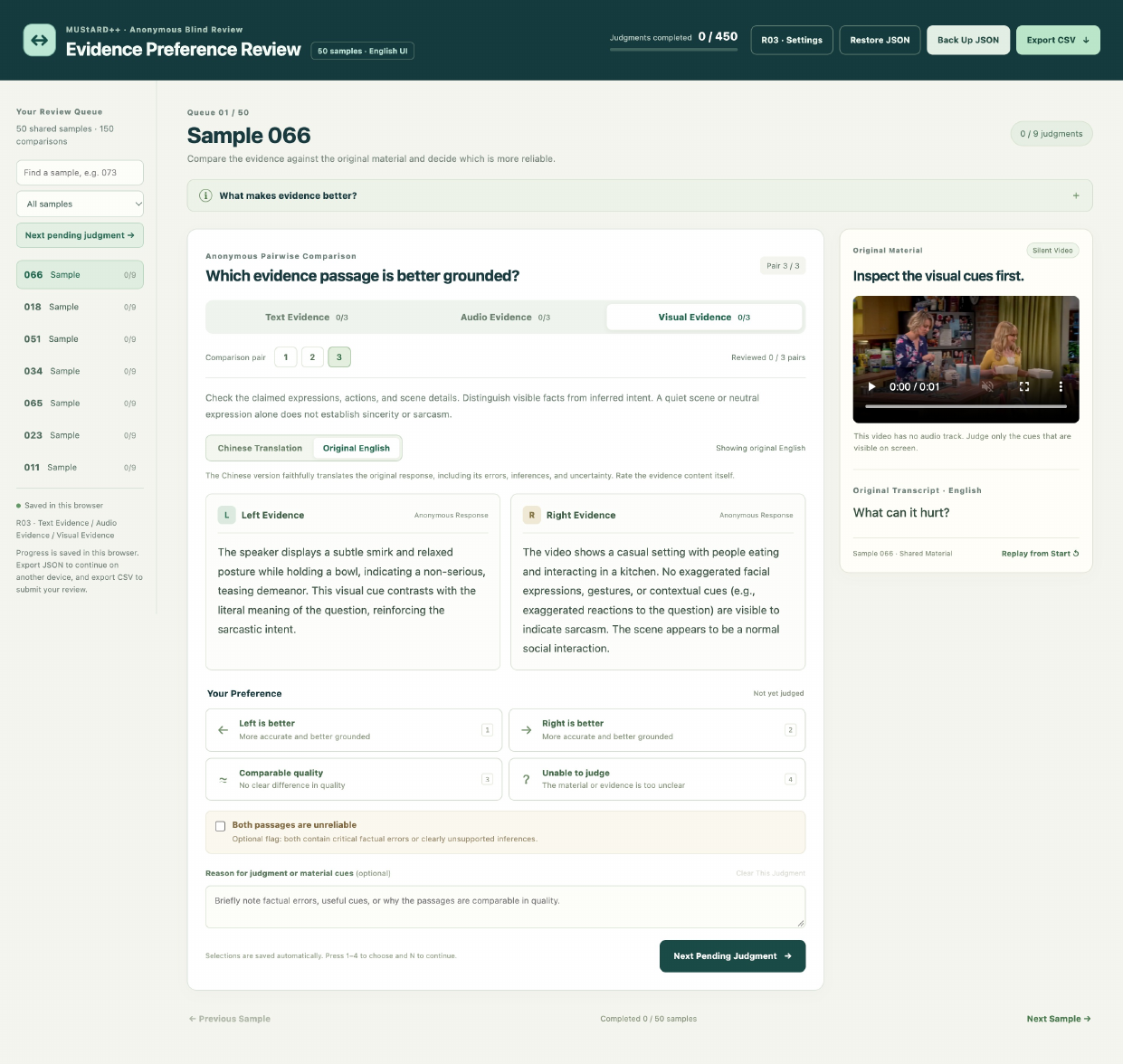}
\caption{Human evaluation interface for visual evidence grounding.}
\label{fig:human_eval_visual}
\end{figure*}

The pairwise human evaluation uses 50 test utterances from each dataset and compares three post-training systems: SFT, SFT+GRPO, and SarcasmReasoner (SFT+GRPO+GenRM). Three PhD researchers independently assess the same model-pair comparisons. For each utterance, the three system pairs are assessed separately for textual, acoustic, and visual evidence grounding. Annotators inspect the corresponding original input and two anonymized generated explanations, select the explanation better supported by that modality, or indicate a tie. Model identities are concealed and the order of the compared outputs is randomized. 
The modality-specific evaluation interfaces are shown in Figs.~\ref{fig:human_eval_text}--\ref{fig:human_eval_visual}.

\noindent\textbf{Inter-annotator agreement.} We quantify agreement across three human annotators' ratings using Fleiss' $\kappa$. Pairwise preferences are coded by the preferred system (SFT, SFT+GRPO, or SarcasmReasoner), with ties treated as a separate category. Table~\ref{tab:human_agreement} reports agreement for each evidence modality, the unweighted mean of the three modality-specific $\kappa$ values, and the overall $\kappa$ calculated by pooling all model-pair and modality judgments within each dataset. The overall agreement is $\kappa=0.647$ on MUStARD++ and $\kappa=0.655$ on MCSD. Modality-specific agreement is highest for textual evidence on MUStARD++ ($\kappa=0.632$) and acoustic evidence on MCSD ($\kappa=0.737$).

% \caption{Inter-annotator agreement measured by Cohen's $\kappa$ for pairwise human preferences. Avg. is the unweighted mean across text, audio, and visual judgments; Overall is computed by pooling all model-pair and modality judgments within each dataset.}

\begin{table}[!htbp]
\centering
\caption{Agreement measured by Fleiss' $\kappa$ across three human annotators. Avg. is the unweighted mean across text, audio, and visual judgments; Overall is computed by pooling all model-pair and modality judgments within each dataset.}
\label{tab:human_agreement}
\small
\setlength{\tabcolsep}{6pt}
\renewcommand{\arraystretch}{1.07}
\begin{tabular*}{\columnwidth}{@{\extracolsep{\fill}}lcc@{}}
\toprule
\textbf{Modality} & \textbf{MUStARD++} & \textbf{MCSD} \\
\midrule
Text & 0.632 & 0.613 \\
Audio & 0.604 & 0.737 \\
Visual & 0.611 & 0.599 \\
\midrule
Avg. & 0.615 & 0.650 \\
Overall & \textbf{0.647} & \textbf{0.655} \\
\bottomrule
\end{tabular*}
\end{table}

\noindent\textbf{Statistical Analysis.}
To account for the non-independence of judgments on the same utterance, we aggregate the three annotators' preferences at the utterance level.
Each judgment is assigned a value of $+1$, $0$, or $-1$, corresponding to a preference for the first system, a tie, or a preference for the second system.
The sign of the sum determines the utterance-level preference, with zero indicating a tie.
We apply two-sided exact sign tests to non-tied utterance-level outcomes and adjust for multiple comparisons using Holm's correction across all 18 tests.

\subsection{Reward Design and Training Configuration}
\label{app:training_setup}

\noindent\textbf{Reward definitions.} Both GRPO variants initialize independently from the same SFT checkpoint and are trained for two epochs. Standard GRPO optimizes accuracy and output format, whereas SarcasmReasoner additionally incorporates a GenRM-based grounding reward. For a sampled trajectory $o_i$ and gold label $c$, the binary accuracy reward is
\begin{equation}
R_{\mathrm{acc}}(o_i)=\mathbb{I}[\hat{c}_i=c],
\end{equation}
where $\hat{c}_i$ is the parsed final prediction; unparseable predictions receive zero reward. The format reward is
\begin{equation}
R_{\mathrm{fmt}}(o_i)=\begin{cases}
1, & \text{if all required elements are present and valid},\\
0, & \text{otherwise.}
\end{cases}
\end{equation}
A valid format contains exactly one nonempty occurrence of each of the five required elements and a permitted final label.

The frozen GenRM predicts scores $\hat{s}_{i,m}\in\{1,2,3\}$, which are normalized to $\{0,0.5,1\}$. The grounding reward is the mean normalized score, gated by prediction correctness and evidence completeness:
\begin{equation}
\begin{aligned}
R_{\mathrm{GenRM}}(x,o_i)&=R_{\mathrm{acc}}(o_i)\,B(o_i)\\
&\quad\times\frac{1}{3}\sum_{m\in\mathcal{M}}\frac{\hat{s}_{i,m}-1}{2},
\end{aligned}
\end{equation}
where $B(o_i)=1$ when the text, audio, and visual evidence elements each occur exactly once and are nonempty, and $B(o_i)=0$ otherwise. This correctness and completeness gating is applied during RL training; the GenRM-based diagnostic scores reported in Table~\ref{tab:genrm_test_reasoning} are the normalized grounding-score averages, not the gated training reward.

\noindent\textbf{Progressive grounding weight.} As described in Section~III-C, the total SarcasmReasoner reward adds $\lambda_g\alpha(p)R_{\mathrm{GenRM}}$ to the accuracy and format terms. We set $\lambda_a=1$ and $\lambda_f=\lambda_g=0.1$. For progress $p\in[0,1]$ within the SarcasmReasoner run, the scheduled multiplier is
\begin{equation}
\alpha(p)=\begin{cases}
0, & p\leq 0.5,\\
\dfrac{p-0.5}{0.1}, & 0.5<p<0.6,\\
1, & p\geq 0.6.
\end{cases}
\end{equation}
Thus, grounding contributes no reward during the first half of SarcasmReasoner training, is introduced linearly over the next 10\% of training, and subsequently reaches its full weight $\lambda_g$. In standard GRPO, the grounding term is disabled ($\lambda_g=0$). Both RL variants use the same policy initialization rather than sequentially training SarcasmReasoner from a completed GRPO checkpoint.

% \noindent\textbf{Checkpoint selection and hyperparameters.} We select the GenRM checkpoint using the mean validation Macro-F1 over its text, audio, and visual score predictions. Figure~\ref{fig:genrm_checkpoint_selection} shows the score-prediction trajectories, and Table~\ref{tab:training_details} summarizes the optimizer, sampling, and model settings.

% \begin{figure*}[t]
% \centering
% \subfloat[MUStARD++.\label{fig:genrm_ckpt_mustard}]{%
%   \includegraphics[width=.485\textwidth,trim=5 5 5 5,clip]{figs/Appendix/genrm_checkpoint_selection_mustard.pdf}}
% \hfill
% \subfloat[MCSD.\label{fig:genrm_ckpt_mcsd}]{%
%   \includegraphics[width=.485\textwidth,trim=5 5 5 5,clip]{figs/Appendix/genrm_checkpoint_selection_mcsd.pdf}}
% \caption{GenRM checkpoint selection on validation data. Curves show
% Macro-F1 for text, audio, and visual score prediction, together with
% their arithmetic mean; the checkpoint maximizing the mean is selected.}
% \label{fig:genrm_checkpoint_selection}
% \end{figure*}

\begin{table}[!htbp]
\centering
\caption{Training hyperparameters. The SFT and RL models use Qwen2.5-Omni-7B; GenRM uses Qwen2.5-Omni-3B. 
% All RL runs are initialized from SFT. 
Dashes denote non-applicable settings.}
\label{tab:training_details}
\footnotesize
\setlength{\tabcolsep}{2pt}
\begin{tabular*}{\columnwidth}{@{\extracolsep{\fill}}lcccc@{}}
\toprule
\textbf{Setting} & \textbf{SFT} & \textbf{GenRM} &
\textbf{GRPO} & \textbf{SarcasmReasoner} \\
\midrule
Model size & 7B & 3B & 7B & 7B \\
Initialization & Base & Base & SFT & SFT \\
Epochs & Varies & 3 & 2 & 2 \\
Learning rate & $10^{-4}$ & $2\cdot10^{-5}$ & $10^{-5}$ & $10^{-5}$ \\
Batch / GPU & 1 & 1 & 1 & 1 \\
Grad. accumulation & 16 & 16 & 8 & 8 \\
LoRA rank / alpha & 8 / 32 & 8 / 32 & 8 / 32 & 8 / 32 \\
Max. seq. length & 16,384 & 16,384 & 16,384 & 16,384 \\
Generations ($G$) & -- & -- & 8 & 8 \\
Max. completion & -- & -- & 1,024 & 1,024 \\
Temperature & -- & -- & 1.0 & 1.0 \\
Top-$p$ & -- & -- & 0.95 & 0.95 \\
KL coefficient $\beta$ & -- & -- & 0.04 & 0.04 \\
\bottomrule
\end{tabular*}
\end{table}

\noindent\textbf{Compute and test-time decoding.} Each training run uses a single NVIDIA H100 GPU. Table~\ref{tab:compute_cost} reports logged GPU-hours to the checkpoints used for the reported results. For test-time evaluation of the post-trained variants, we generate one response per instance using deterministic greedy decoding with temperature 0 and at most 4,096 new tokens, without sampling or ensembling, and parse the final label from the \texttt{<answer>} section.

\begin{table}[!htbp]
\centering
\caption{GPU-hours required to reach the checkpoints used for the reported results. Every run uses one NVIDIA H100, so GPU-hours equal logged wall-clock hours. Values cover the listed training segment and exclude upstream checkpoint construction, teacher trajectory generation, judge annotation, and test-time evaluation.}
\label{tab:compute_cost}
\small
\setlength{\tabcolsep}{4pt}
\begin{tabular*}{\columnwidth}{@{\extracolsep{\fill}}llrr@{}}
\toprule
\textbf{Training segment} & \textbf{Model} & \textbf{MUStARD++} & \textbf{MCSD} \\
\midrule
Diverse-8 SFT & Qwen2.5-Omni-7B & 3.1 & 11.9 \\
GenRM SFT & Qwen2.5-Omni-3B & 2.8 & 6.6 \\
Standard GRPO & Qwen2.5-Omni-7B & 12.1 & 46.1 \\
SarcasmReasoner & Qwen2.5-Omni-7B & 13.0 & 50.2 \\
\bottomrule
\end{tabular*}
\end{table}

\subsection{Visual-Evidence Length and No-Cue Statements}
\label{app:visual_evidence_analysis}

To examine whether improvements in visual grounding could partly reflect changes in response style rather than better evidence grounding, we analyze the visual-evidence sections generated by Diverse-8 SFT, standard GRPO, and SarcasmReasoner on the full test sets of both datasets. We consider two descriptive measures: (1) the mean number of words in each visual-evidence section, and (2) the proportion of sections explicitly stating that no relevant visual cue supports a sarcastic or ironic interpretation. The latter captures the frequency of such statements, without assessing whether the claimed absence of visual evidence is correct.

\begin{table}[t]
\centering
\caption{Response-style statistics for generated visual-evidence sections on the full test sets. Mean words denotes the average section length; No cue denotes the percentage of sections explicitly reporting an absence of visual evidence supporting sarcasm or irony.}
\label{tab:visual_evidence_style}
\small
\setlength{\tabcolsep}{5pt}
\renewcommand{\arraystretch}{1.08}
\begin{tabular*}{\columnwidth}{@{\extracolsep{\fill}}lrr@{}}
\toprule
\textbf{System} & \textbf{Mean words} & \textbf{No cue (\%)} \\
\midrule
\multicolumn{3}{@{}l}{\emph{MUStARD++} ($N=181$)} \\
SFT (Diverse-8) & 44.04 & 59.67 \\
GRPO            & 44.13 & 58.01 \\
SarcasmReasoner & 49.25 & 77.90 \\
\addlinespace[3pt]
\multicolumn{3}{@{}l}{\emph{MCSD} ($N=406$)} \\
SFT (Diverse-8) & 55.00 & 69.46 \\
GRPO            & 55.07 & 63.05 \\
SarcasmReasoner & 49.83 & 47.04 \\
\bottomrule
\end{tabular*}
\end{table}

As shown in Table~\ref{tab:visual_evidence_style}, the response-style changes differ across datasets. On MUStARD++, SarcasmReasoner generates longer visual-evidence sections than standard GRPO (49.25 vs.\ 44.13 words), while also producing more explicit no-cue statements (77.90\% vs.\ 58.01\%). This suggests that a more cautious treatment of visual evidence may partly contribute to its higher grounding scores on this dataset. In contrast, on MCSD, SarcasmReasoner generates shorter sections (49.83 vs.\ 55.07 words) but fewer no-cue statements (47.04\% vs.\ 63.05\%). Thus, the improvements in visual-grounding scores are not consistently associated with shorter or more cautious descriptions across datasets. Nevertheless, these descriptive measures cannot establish whether individual visual claims are correctly grounded, and response-style effects cannot be fully ruled out.

\subsection{Prompt Templates}
\label{app:prompts}

\subsubsection{Teacher Reasoning Prompt}
\label{app:teacher_prompt}
The teacher is instructed to examine linguistic, acoustic, and visual information separately before integrating the evidence into a final sarcasm decision. The complete prompt is reproduced in Table~\ref{tab:sarcasm_reasoning_prompt}.

\begin{table*}[!t]
\centering
\begin{tcolorbox}[
  width=\textwidth,
  colback=white,
  colframe=black!40,
  boxrule=0.5pt,
  arc=0.8mm,
  left=2mm,right=2mm,top=1.4mm,bottom=1.4mm,
  title=\textbf{Sarcasm Reasoning Prompt},
  fonttitle=\small
]
\begin{Verbatim}[fontsize=\footnotesize,breaklines=true,breakanywhere=true,breaksymbolleft={},breaksymbolright={}]
Analyze the provided utterance, audio, and video and determine whether the speaker is being sarcastic.

Evaluate the modalities independently before integrating them:

1. Text: Identify only textual or semantic evidence actually present in the transcript.
2. Audio: Identify only audible prosodic or paralinguistic evidence clearly present in the audio.
3. Visual: Identify only visible facial, gestural, or contextual evidence clearly present in the video.
4. Integration: Explain how the available evidence across modalities supports the final interpretation.

Do not force every modality to provide evidence.
If a modality contains no clear relevant cue, state this explicitly.

Do not treat humor, exaggeration, joking, audience laughter, or a comedy-show setting alone as evidence of sarcasm.

Distinguish sarcasm from non-sarcastic humor, exaggeration, joking, storytelling, and straightforward criticism.

Sarcasm does not require the literal meaning to be the exact opposite of the intended meaning. Ironic framing, pretended stance, evaluative incongruity, or deliberately incongruous wording may also provide evidence of sarcasm.

Do not infer an audio or visual cue from the general scene, genre, or expected context.
For example, do not claim audience laughter unless it is actually audible, and do not infer a facial expression or gesture unless it is clearly visible.

Do not speculate about information that is not observable.
Do not assume that a stated reason, attitude, or intention is insincere unless there is observable evidence supporting that interpretation.

Write the reasoning in English.

Return exactly:

<text_evidence>
...
</text_evidence>

<audio_evidence>
...
</audio_evidence>

<visual_evidence>
...
</visual_evidence>

<integration>
...
</integration>

<answer>
...
</answer>

The answer must be exactly either "Sarcasm" or "Non-Sarcasm".
Keep the final reasoning concise and evidence-grounded.
\end{Verbatim}
\end{tcolorbox}
\caption{Prompt used to elicit structured, evidence-grounded multimodal sarcasm reasoning from the teacher model.}
\label{tab:sarcasm_reasoning_prompt}
\end{table*}

\subsubsection{Evidence-Quality Judge Prompt}
\label{app:judge_prompt}
The quality judges evaluate whether each candidate evidence section is
supported by its corresponding input modality, without using an
independently inferred sarcasm label as a substitute for grounding.
Table~\ref{tab:judge_prompt} reproduces the shared scoring rubric used
by the modality-specific judges. The full model-specific prompts,
including additional safeguards, will be provided with the code in an online repository.

\begin{table*}[!t]
\centering
\begin{tcolorbox}[
  width=\textwidth,
  colback=white,
  colframe=black!40,
  boxrule=0.5pt,
  arc=0.8mm,
  left=2mm,right=2mm,top=1.4mm,bottom=1.4mm,
  title=\textbf{Evidence-Quality Judge Prompt},
  fonttitle=\small
]
\begin{Verbatim}[fontsize=\footnotesize,breaklines=true,breakanywhere=true,breaksymbolleft={},breaksymbolright={}]
Evaluate evidence-grounded reasoning in the TEXT, AUDIO, and VISUAL candidate sections.

For each section, judge both whether its important observations are supported by the corresponding modality and whether its interpretations follow at the stated level of certainty.

Evaluate the argument as written. Do not supply missing premises to repair it. Do not independently classify sarcasm or reward agreement with an implied label.

Evidence boundaries:

TEXT:
Use the transcript and ordinary linguistic understanding only.

AUDIO:
Use the audio, with the transcript as linguistic context.

VISUAL:
Use the ordered video frames, with the transcript as linguistic context.

Evidence from one modality must not be used to rescue unsupported claims in another modality.

Scoring:

1 = UNRELIABLE
No meaningful evidence basis remains, OR the main argument depends on a material misreading, unsupported premise, or invalid inference. Correct quotations or incidental true observations do not rescue a failed core argument.

2 = PARTIALLY RELIABLE
A substantive supported account or useful argument remains, BUT an important observation or inference is unsupported, contradicted, or overstated. The section is reliable only in part, rather than fundamentally unsupported.

3 = RELIABLE
The section has a meaningful evidence basis. All important observations are supported and all important interpretations are reasonable at the stated certainty. No material factual or reasoning problem remains.

Return exactly:
{"text":X,"audio":Y,"visual":Z}

X, Y, and Z must each be an integer from 1 to 3.
\end{Verbatim}
\end{tcolorbox}
\caption{Condensed common rubric for modality-specific evidence-quality judges. The full modality-specific prompts will be provided with the code in an online repository.}
\label{tab:judge_prompt}
\end{table*}

\end{document}